\documentclass[12pt]{article}
\usepackage{amsmath}
\usepackage{amssymb}
\usepackage{amsfonts}
\usepackage{amscd}
\usepackage{delarray}
\usepackage{graphicx}

\usepackage{amsmath,amsmath,amsfonts,amssymb,color,bbm}

\usepackage{graphicx}

\def\Symp#1,#2,#3,#4.{\left[\left(\begin{array}{c}#1\\#2\end{array}\right),\left(\begin{array}{c}#3\\#4\end{array}\right)\right]}
\def\Vec#1,#2.{\left(\!\begin{array}{c}#1\\#2\end{array}\!\right)}
\def\vec#1,#2.{{#1\choose{#2}}}
\def\ket#1.{|#1\rangle}

\def\bra#1.{\langle#1|}
\def\braket#1,#2.{\langle#1|#2\rangle}
\newcommand{\beq}{\begin{equation}}
\newcommand{\eeq}{\end{equation}}
\newcommand{\beqa}{\begin{eqnarray}}
\newcommand{\eeqa}{\end{eqnarray}}

\begin{document}

\title{Crucial tests of the existence of a Time Operator.}
\date{}
\author{}
\maketitle
\vglue -1.8truecm
\centerline{Thomas Durt\footnote{TONA Vrije Universiteit Brussel, Pleinlaan 2, B-1050
Brussels, Belgium. \texttt{email:
thomdurt@vub.ac.be}} }

%



\begin{abstract}
 In the present paper we show that the {\it Temporal Wave Function} approach of the decay process, which is a multicomponent version of the {\it Time Operator} approach leads to new, non-standard, predictions concerning the statistical properties of decay time distributions of single kaons and entangled pairs of mesons. These results suggest crucial experimental tests for the existence of a Time Operator for the decay process to be realized in High Energy Physics or Quantum Optics.

\end{abstract}

\maketitle

\section*{Introduction}
The status of Time in the Quantum Theory is still a controversial
subject \cite{booktime}. For instance, it is well known that time-energy uncertainty relations have not the same theoretical status as position-momentum uncertainty relations. 

One of the reasons thereof is that in the first quantized version of quantum mechanics {\it \`a la Schr\"odinger}, time is an external parameter, and the predictions of the theory concern outcomes of measurements (for instance of position) realized at a fixed time $t$. Therefore, although standard quantum mechanics allows us
in principle to bring an unambiguous answer to the question {\it Is
a certain particle located inside a given space region at time t ?}, it does not allow \cite{mielnik} to answer unambiguously to the question {\it
At what time will the particle enter this region?}\footnote{One could
equivalently formulate this question in the form  {\it What is the
probability that a particle hits a screen during a given time interval?} or {\it At what time
will the particle hit the screen?}, which justifies why the problem
of deriving a temporal distribution from Schr\"odinger's wave
function (instead of a spatial distribution) is often refered to in
the literature as the so-called {\it screen problem} \cite{mielnik} or arrival-time problem \cite{cushing}.} 

Second quantization does not help us to solve the problem. Actually, even localisation in space becomes ill-defined in Quantum Field Theory \cite{wignernewton} so that,  in a sense, the situation is worse in that case. The conceptual difficulties that characterize the attempts to formulate the non-relativistic spontaneous localisation theory of Ghirardi Rimini and Weber in a Lorentz covariant fashion are another illustration of the problem \cite{BG03}. More generally, the EPR paradox \cite{EPR} and the issues related to quantum non-locality and Bell inequalitites \cite{Bell} also emphasize how difficult it is to find a relativistic description of individual quantum measurement processes.

A main difference between space and time in this context is that in the first quantization approach the position of a particle is assumed to be distributed, at a given instant of time, randomly in space according to the $|\psi|^2$ distribution. Beside that, particles are assumed to be ''solid'' objects that exist forever, so that it does not make sense to associate a Temporal Wave Function to their description (they exist at any time with constant probability equal to 1). 

Now, in the case of decaying particles, the problem is other: if one prepares at time 0 a particle in an unstable state, the particle is likely to decay after a time comparable to the lifetime of the excited unstable state, and the exact time at which the decay occurs is a random variable, that is distributed according to a well-defined statistical distribution. Besides, it has been shown empirically that the uncertainty on the energy of the initial state and the uncertainty on the time at which the particle decays obey an Heisenberg-like complementarity relation \cite{busch} (see also appendix 1 concerning this point). This suggests that one can, in analogy with what is done regarding position in the first quantization procedure, associate a wave function to the ``observable'' time of decay, on the basis of the substitutions 

i) Position $x$ $\leftrightarrow$ Time of Decay $t$

ii) Spatial Wave Function $\psi(x)$  $\leftrightarrow$ Temporal Wave Function $\psi(t)$

iii) Probability to find the particle between $x$ and $x+dx$=  $|\psi(x)|^2dx$ $\leftrightarrow$ Probability that the particle decays between $t$ and $t+dt$=  $|\psi(t)|^2dt$.

This is the essence of the wave-function approach which is, as we have shown in ref.\cite{09superop} equivalent to the Time Operator approach (in the case of exponential decay). Up to now, the controversial question of the existence of a Time Operator that could describe the decay process remained confined to the realm of theory and speculation\footnote{We also discuss the status of the Time Operator in appendix 1.}. In a previous paper we proposed, in the framework of the wave-function approach a simple quantum model of the
kaons decay \cite{09temporal} and of the CP-violation process that has been observed in 1964 during the Cronin and Fitch experiment \cite{christ}. The aim of the present paper is to analyse in depth the differences between the Time Operator approach and the standard approach and to propose experimental tests aimed at discriminating both approaches. 

The paper is structured as follows. In the first section, we present the standard and non-standard (Temporal Wave Function or Time Operator that we shall most often label TWF-O) approaches to the decay process. In the second one we recall basic concepts regarding kaon phenomenology. In the third one we we consider a situation for which standard and non-standard approaches lead to similar predictions and allow us to model a dichotomic measurement as a continuous in time process. In particular we show how, in the TWF-O approach the standard temporal statistics of detector clicks can be simulated with the help of an {\it effective} two-component wave function, based on an analogy with a spatial stationary spinorial distribution. In the fourth section we apply both approaches in order to model the Cronin and Fitch \cite{christ} experiment, that was the first experiment to reveal CP violation. In the fifth section we show how a fine study of the statistics of decay times measured during the Cronin and Fitch experiment would make it possible to discriminate the standard and non-standard approaches. In the sixth section we consider two infinite families of entangled bipartite mesonic states. Each of them comprises two so-called {\it Bell} states.  We show that both approaches lead to the same predictions when the entangled state belongs to the class that comprises the {\it singlet} state (that has been realized experimentally in the past). On the contrary, both predictions differ when the second entanglement-class is considered, which also opens the way to crucial experiments. The last section is devoted to the conclusion and to discussions.

\section{Discrepancy between the wave-function approach and the standard approach.}
\subsection{Standard approach to the decay processes.}

Although
there is no unambiguous {\it recipe} for deriving temporal probability
distributions in the quantum theory (see e.g. introduction of ref.\cite{booktime} and references therein), in the majority of standard
approaches to the decay process (among which the celebrated Wigner-Weisskopf approach \cite{WW}), metastable states or resonances are
formally characterized by a complex energy, of which the real part
is proportional to the mass of the particle and of which the
imaginary part is proportional to the inverse of its lifetime.

Let us denote $P_s(t)$ the ``\emph{integrated
survival probability}''   from an initial time 0 up to $t$.

In the case of a state with a complex energy, the integrated
survival probability $P_s(t)$ obeys 

\begin{equation}P_s(t)=\psi^*(t)\psi(t)/\psi^*(0)\psi(0).\label{exp}\end{equation}

 Indeed, setting (for positive times $t$)
\begin{equation}\psi(t)=\psi(0) e^{-\mathrm{i}E t}=\psi(0) e^{-\mathrm{i}(m-\frac{\mathrm{i}}{2}\Gamma) t},\label{gamow}\end{equation} we
get \cite{gamow}, as a consequence of equation (\ref{exp}),
\begin{equation}P_s(t)=e^{-{ t\over \tau}}\end{equation}
 where the lifetime $\tau$  is  equal to the inverse of
the so-called decay rate $\Gamma$: $\Gamma={1\over \tau}$.

In the case that $\psi(t)$ is a multicomponent function of time, the generalisation of expression (\ref{exp}) reads

\begin{equation}P_s(t)=\|\psi(t)^2\|/\|\psi(0)^2\|\label{exp2},\end{equation}

with $\| \psi\|^2=\sum_{i=1}^N\psi^*_i(t)\psi_i(t)$, with for instance $N=2$ in the case of the quasi-spin picture that is often used to describe mesons such as kaons or $B$ and $D$ particles.

The equation (\ref{exp}) is widely used in the treatment of decaying systems, and its validity is also assumed in the case of non-exponential decay processes, for instance in the case that the state of the particle coherently oscillates between two states described by different complex energies (and thus different masses and life times). This occurs in kaon phenomenology, which makes that kaons are   fascinating quantum systems: a neutral kaon can {\emph ''exist in a coherent superposition of two different particle-states''} ($\mathrm{K}^0$ and $\overline{\mathrm{K}}^0$) that differ among others by their disintegration products in matter, and by the ''{\emph Strangeness}'' quantum number $S$. Besides, freely propagating kaons can also be expressed as superpositions of states that are characterized by radically different life times in vacuum (the so-called Short-lived and Long-lived states) and are in first approximation characterized by different CP eigenvalues.

Let us denote $p_d(t)$ the (temporal) ``\emph{density of probability}'' of decay or \emph{decay rate}
which is equal to minus the time derivative of $P_s(t)$:
\begin{equation}\label{pd}p_d(t)=d(1-P_s(t))/dt=-dP_s(t)/dt.\end{equation} 

The (temporal) density of probability of decay $p_d(t)$ or {\it probability density function} (pdf) of decay is  not measured in the same way as the integrated survival probability as we shall discuss in a forthcoming section (\ref{stand}).  In the case of an exponential decay
nevertheless this distinction is often overlooked without
prejudicial consequences because $P_s(t)$ is {\it proportional} to
$p_d(t)$. 

Indeed, deriving both sides of the expression (\ref{exp}) relatively to time, we obtain

\begin{equation}p_d(t)=-dP_s(t)/dt=\Gamma\cdot e^{-\Gamma t}.=p_d(0)\cdot e^{-\Gamma t}.=p_d(0)\cdot P_s(t),\label{exprate}\end{equation}so that the pdf and the (integrated) survival probability differ only by a global constant.
In the majority of experiments, what is measured is a relative detection rate so that a global factor is easily absorbed in the calibration of the detector(s).

\subsection{Temporal Wave Function (Time Operator) approach (TWF-O) to the decay process.}

One can also, still in the case of exponential decays, express the decay rate in function of the instantaneous modulus square of the ''Gamow'' wave function (\ref{gamow}) as follows
\begin{equation}p_d(t)=-dP_s(t)/dt=p_d(0){\psi^*(t)\psi(t)\over \psi^*(0)\psi(0)}.\label{exprategen}\end{equation}

The multicomponent version of (\ref{exprategen}) reads

\begin{equation}p_d(t)=-dP_s(t)/dt=p_d(0){\|\psi(t)^2\|\over \|\psi(0)^2\|}.\label{exprategen2}\end{equation} with $\| \psi\|^2=\sum_{i=1}^N\psi^*_i(t)\psi_i(t)$.

It is worth noting that the expression (\ref{exprategen}) is reminiscent of the substition (iii) mentioned in the introduction and that, excepted in the case of purely exponential decay-time distributions, it radically differs from the standard approach in which it is rather the expression (\ref{exp}) (or \ref{exp2}) that is considered to establish a link between the wave function of the decaying particle and the distribution of decay times (\cite{girad}, see also appendix 1 in relation with this question).

In the next sections we shall compare the predictions made on the basis of the ``standard'' expression (\ref{exp}) and those that relie on the ``alternative'' (TWF-O) expression (\ref{exprategen}). In particular we shall focus on a model that we elaborated in the past \cite{09temporal} in which instead of the conventional {\it ''quasi-spin''} analogy we developed a {\it ''quasi-spin Temporal Wave Function analogy''} for the decay process. In ref.\cite{09temporal} we have shown that the latter, non-standard, approach, in the case of single kaon experiments, leads to similar predictions as the former, standard, approach for what concerns CP-violation. In the present paper we show how to conceive crucial experiments for which the standard and non-standard approaches lead to different predictions which would allow us to discriminate them experimentally. 

The reader could legitimately wonder why one should question the standard approach that appeared to be so useful in the past. There are several reasons that motivated our approach. 

Firstly we realized from the study of the Time (Super) Operator that other, non-standard, approaches to the problem of the distribution of time decays are possible and that there exist crucial experiments that would allow in last resort to ''discriminate'' the different approaches \cite{09superop}. 

Secondly we noticed that in the literature some confusion exists regarding how one should interpret the wave function in the case of decaying processes. Depending on the reference, the priority is given sometimes to the \emph{integrated survival probability} $P_s(t)$ and other times to the (temporal) ``\emph{density of probability}'' of decay $p_d(t)$. As we shall discuss in the paper (section \ref{stand}), it is not very clear what is exactly measured during experimental tests of CP-violation. It is absolutely necessary to clarify this ambiguity experimentally and theoretically as well because its answer conditions our modellisation of the phenomenon. 

Thirdly, we consider that the expressions (\ref{exp},\ref{exp2}) have still, to some extent, the status of a ''recipe''\footnote{\label{monotonous}As it was noted by Misra and Sudarshan and several others \cite{misrasud,chiuchiu,arrival}, a survival probability is {\it per se} a monotonously decreasing function of time which imposes severe constraints, together with the expressions (\ref{exp},\ref{exp2}), onto the function $\psi(t)$. In the presence of Rabi oscillations these constraints are not always fulfilled (see also appendix 1). On the contrary, when the validity of the expressions (\ref{exprategen},\ref{exprategen2}) is postulated, the sign of the derivative of the survival probability is {\it de facto} negative.}, of which the validity could be questioned and even challenged experimentally as we shall show in the present paper. 

Finally, the idea that a Temporal Wave Function would make sense is appealing in itself and could shed a new light on the role played by Time on the Quantum Theory. After all, we are here in a {\emph ''no man's land''} where the standard quantum theory {\it \`a la Schr\"odinger} is not necessarily valid \cite{ghirardiEPR}, and where intuition remains the best guide. One of our main motivations in writing this paper is also to clarify foundational issues related to the temporal wave function approach and to the possible existence of a Time Operator.

\section{Kaon phenomenology.} Let us first recall some basic experimental facts. Kaons are mesons that were discovered in the forties during the study of cosmic rays.
They are produced by collision processes in nuclear reactions during
which the strong interactions dominate. They appear in pairs
$\mathrm{K}^{0}$, $\overline{\mathrm{K}}^{0}$ \cite{perkins,hokim}.

In the standard approach, kaons are treated like two-level systems in analogy with spin 1/2 particles or photonic polarisations. This is the so-called {\it ''quasi-spin''} picture introduced by Lee and Wu \cite{Lee} and Lipkin \cite{Lipkin}. In the overview paper  \cite{bertlmann} of Bertlmann (Entanglement, Bell inequalities and Decoherence in Particle Physics), the analogy with photonic polarisations is explained very clearly. We advice the interested reader to consult this useful reference.

The $\mathrm{K}$ mesons are eigenstates of the parity operator $P$:
$ P|\mathrm{K}^0\rangle=- |\mathrm{K}^0\rangle$, and $
P|\overline{\mathrm{K}}^0\rangle=- |\overline{\mathrm{K}}^0\rangle$.
$\mathrm{K}^0$ and $\overline{\mathrm{K}}^0$ are charge conjugate to
each other  $ C|\mathrm{K}^0\rangle=
|\overline{\mathrm{K}}^0\rangle$, and $
C|\overline{\mathrm{K}}^0\rangle= |\mathrm{K}^0\rangle$. We get
thus,
$CP|\mathrm{K}^0\rangle= -|\overline{\mathrm{K}}^0\rangle,$ and $
 CP|\overline{\mathrm{K}}^0\rangle=
-|\mathrm{K}^0\rangle.$
Clearly $|\mathrm{K}^0\rangle$ and $|\overline{\mathrm{K}}^0\rangle$
are not $CP$-eigenstates, but the  following combinations
\begin{equation}\label{kk1}
\nonumber|\mathrm{K}_1\rangle=\frac{1}{\sqrt{2}}\big{(}|\mathrm{K}^0\rangle
-|\overline{\mathrm{K}}^0\rangle\big{)},~~~
|\mathrm{K}_2\rangle=\frac{1}{\sqrt{2}}\big{(}|\mathrm{K}^0\rangle
+|\overline{\mathrm{K}}^0\rangle\big{)},
\end{equation}
are $CP$-eigenstates:
$CP|\mathrm{K}_1\rangle=|\mathrm{K}_1\rangle$ and $
CP|\mathrm{K}_2\rangle=-|\mathrm{K}_2\rangle. $ 
In the
absence of matter, kaons disintegrate through weak interactions
\cite{hokim}.

Actually, $\mathrm{K}^0$ and $\overline{\mathrm{K}}^0$
are distinguished by their mode of \emph{production} while $\mathrm{K}_1$
and $\mathrm{K}_2$ are in first approximation (in absence of
$CP$-violation) the decay modes of the kaons so that the weak disintegration process distinguishes the
$\mathrm{K}_{1}$ states which decay only into ``$2\pi$'' while the
$\mathrm{K}_{2}$ states decay into ``$3\pi, \pi e \nu, ...$''
\cite{leebook}. The lifetime of the $\mathrm{K}_{1}$ kaon is short
($\tau_{S}\approx 8.92\times10^{-11}~^\mathrm{s}$), while the
lifetime of the $\mathrm{K}_{2}$ kaon is quite longer
($\tau_{L}\approx 5.17\times10^{-8}~^\mathrm{s}$) \cite{perkins}.

$CP$-\emph{violation}  was discovered by Christenson {\it et al.} \cite{christ} in the so-called Cronin and Fitch experiment. $CP$-violation is revealed by the fact that the long-lived kaon can
also decay to ``$2\pi''$. Then, the $CP$ symmetry is slightly
violated (by a factor of $10^{-3}$) by weak interactions so that the
$CP$ eigenstates $\mathrm{K}_1$ and $\mathrm{K}_2$ are not exact
eigenstates of the decay interaction. Those exact states are
characterized by lifetimes that are in a ratio of the order of
$10^{-3}$, so that they are called the short-lived state
($\mathrm{K}_S$) and long-lived state ($\mathrm{K}_L$). They can be
expressed as coherent superpositions of the $\mathrm{K}_1$ and
$\mathrm{K}_2$ states through
\begin{eqnarray}\label{kk2}
\nonumber|\mathrm{K}_L\rangle=\frac{1}{\sqrt{1+|\epsilon|^2}}\big{[}
\epsilon
~|\mathrm{K}_1\rangle + |\mathrm{K}_2\rangle \big{]},\\
|\mathrm{K}_S\rangle=\frac{1}{\sqrt{1+|\epsilon|^2}}\big{[}
|\mathrm{K}_1\rangle +\epsilon ~ |\mathrm{K}_2\rangle \big{]},
\end{eqnarray}
where $\epsilon$ is a complex $CP$-violation parameter,
$|\epsilon|\ll1$.
$\mathrm{K}_L$ and $\mathrm{K}_S$ are the eigenstates of the
Hamiltonian for the mass-decay matrix \cite{hokim,leebook}: $ H=M-\frac{\mathrm{i}}{2}\Gamma$
where $M$ and $\Gamma$ are individually Hermitian since they
correspond to observables (mass and lifetime).

The corresponding evolution equation reads in the CP-eigenbasis ($|\mathrm{K}_1\rangle$,$|{\mathrm{K}}_2\rangle$)

\begin{equation}
\mathrm{i}\frac{\partial }{\partial t} \left(\begin{array}{c}
\psi_1(t)\\ \psi_2(t)
\end{array}\right)=
\left(\begin{array}{cc}M_{11}- \mathrm{i} {\Gamma_{11}\over 2} & \ M_{12}-\mathrm{i}  {\Gamma_{12}\over 2} \\
 \ M_{21}- \mathrm{i} {\Gamma_{21}\over 2} & M_{22}- \mathrm{i} {\Gamma_{22}\over 2}\end{array}\right)
\left(\begin{array}{c} \psi_1(t)\\\psi_2(t)
\end{array}\right),\label{fh1}
\end{equation}

 In the practice, the violation is small which means that at the zeroth order approximation one can neglect the off-diagonal factors in the mass decay matrix. As we shall show in the next section, at this order of approximation all approaches (the survival probability approach (\ref{exp}) as well as the wave function approach (\ref{exprategen})) lead to similar predictions. 
 
\section{Comparison of the Standard approach and the Temporal Wave Function approach (TWF-O) in absence of CP violation.\label{sexion}}
\subsection{Standard approach.\label{standard}}  In absence of CP violation, the evolution equation of the quasi-spinor is 
\begin{equation}
\mathrm{i}\frac{\partial }{\partial t} \left(\begin{array}{c}
\psi_1(t)\\ \psi_2(t)
\end{array}\right)=
\left(\begin{array}{cc}M_{S}- \mathrm{i} {\Gamma_{S}\over 2} & 0\\
 0 & M_{L}- \mathrm{i} {\Gamma_{L}\over 2}\end{array}\right)
\left(\begin{array}{c} \psi_1(t)\\\psi_2(t)
\end{array}\right),\label{fh1}
\end{equation}with $M_S=M_{11}, M_L=M_{22}, \Gamma_S=\Gamma_{11}, \Gamma_L=\Gamma_{22}$.
This is equivalent, up to relabellings, to the equation that describes the distribution of a monochromatic transverse electric field (of pulsation $\omega$) in a birefringent guide. Each component would correspond to the projection of the electric field along one of the two (orthogonal) major-axes of the fiber that are characterized by two different velocities of propagation $c_S$ and $c_L$ (with $\omega/c_{S(L)}\leftrightarrow M_{S(L)}$ and $x\leftrightarrow t$) and two absorption coefficients $\Gamma_{S}$ and $\Gamma_{L}$. In such a case, both propagation channels {\it ''don't talk to each other''} and are fully independent. The solution is a superposition of two exponentially damped signals:

\begin{equation}
\left(\begin{array}{c}
|\psi_1(t)|^2\\ |\psi_2(t)|^2
\end{array}\right)=
\left(\begin{array}{c} e^{- \Gamma_{S}t}|\psi_1(0)|^2\\ e^{- \Gamma_{L}t}|\psi_2(0)|^2
\end{array}\right),\label{fh2}
\end{equation}

As for the case of a single exponential decay process, there is no ambiguity here, the \emph{integrated
survival probability} of the Short (Long) component $P_s^{1(2)}(t)$ from an initial time 0 up to $t$ is equal to $e^{- \Gamma_{S(L)}t}$ and satisfies (\ref{exp}) and (\ref{exp2}) as well:

\begin{equation}
\left(\begin{array}{c}
P_s^1(t)\\ P_s^2(t)
\end{array}\right)=
\left(\begin{array}{c} P_s^1(0)e^{- \Gamma_{S}t}\\ P_s^2(0)e^{- \Gamma_{L}t}
\end{array}\right),\label{fh3}
\end{equation}

 while the corresponding (temporal) \emph{density of probability} of decay or \emph{decay rate} $p_d^{1(2)}(t)=-dP^{1(2)}_s(t)/dt$ satisfies  
 \begin{equation}
\left(\begin{array}{c}
p_{d}^1(t)\\ p_{d}^2(t)
\end{array}\right)=
\left(\begin{array}{c}P_s^1(0)\Gamma_{S} e^{- \Gamma_{S}t}\\ P_s^2(0)\Gamma_{L}e^{- \Gamma_{L}t}
\end{array}\right),\label{fh4}
\end{equation}
in accordance with (\ref{exprategen},\ref{exprategen2}). The probability that the particle disintegrates in the CP=+1 (CP=-1) sector between time $t$ and time $t+dt$ is then equal to $P^{S(L)}_s(0)\Gamma_{S(L)}e^{- \Gamma_{S(L)}t}\cdot dt$.

\subsection{Continuous (passive) versus instantaneous (active) measurement.\label{contmeas}}
It is often assumed that quantum measurements occur somewhat {\emph ''out of Time''}. We have in mind here the famous instantaneous collapse of the wave function that is likely to occur during a measurement process, and that led to a celebrated series of paradoxes (EPR, Bell \cite{EPR,Bell} and so on). Of course, each measurement is a physical process that can never be infinitely short. Therefore the picture according to which measurements occur {\emph ''out of Time''} ought, in our view, to be considered as a limit case: ``{\it Instantaneous Measurements''} are seen by us as an idealisation of {\bf very short} {\it ``Continuous Measurements''}.

In the case that the two damping factors $\Gamma_{S}$ and $\Gamma_{L}$ differ by several order of magnitudes (what is the case for what concerns freely propagating kaons for which $\tau_L\approx 1000\cdot\tau_S$), the quasi-spin model developed in the previous sections presents an interesting feature: after a time intermediate between $\tau_S$ and $\tau_L$, the Short component becomes exponentially small and only the Long component survives. 

In such a case, our model provides a modelisation of the collapse process that is more realistic than its instantaneous counterpart in the sense that it describes a continuous, non-instantaneous measurement process.

This also constitutes a better fit to what really happens for instance in a photonic polarising filter, or when kaons are sent through massive matter in order to measure $S$ eigenvalues.

 When a photon enters a polarising filter for instance, the polarisation is not instantaneously filtered, but it is necessary that light interacts for a while with the polarising medium of which the filter is made before an effective collapse occurs. It is only in the ideal and irrealistic limit of an infinitely thin polariser that the filtering becomes instantaneous \cite{papaliolos}.
 
 Similarly, when a kaon is sent through matter, the $S$ eigenstates $|\mathrm{K}^0\rangle$ and $|\overline{\mathrm{K}}^0\rangle$ are characterized by different absorption coefficients, so that one of them is preferentially ``produced'', but this filtering\footnote{Actually, as is explained in 
\cite{bertlmann}, the choice of the basis of measurement is experimentally limited in the case of kaons (for instance, a decay in the presence of matter is dominated by the strong interaction Hamiltonian that is diagonalized by the $S$-diagonal ($\mathrm{K}^{0}$, $\overline{\mathrm{K}}^{0}$) basis while in the vacuum weak interaction dominates and the proper basis is, in first approximation, the CP commuting basis  ($\mathrm{K}_1$, $\mathrm{K}_2$)). } is far from being instantaneous (the efficiency to induce a kaon-nucleon interaction at a given time goes to 1 only in the limit of ultrarelativistic kaons and infinitely dense materials \cite{bramon}).

    In the practice, many measurements realized in the framework of $CP$ phenomenology \cite{christ,frascati}, consist of passively measuring the populations of specific decay products, a process which can be formalized as a continuous measurement of the type considered here above. 
    
    This doesn't mean that instantaneous or nearly instantaneous measurements are forbidden in principle. For instance instead of ``passively'' watching the decay-products of a freely propagating kaon decays during a time interval [$0, \delta \tau$=4.8
    $\cdot \tau_S$] , which constitutes, neglecting CP-violation effects, a $K_1$ versus $K_2$ discrimination characterized by a misidentification probability smaller than one percent \cite{bramon}, an experimentator could instead ``actively'' choose to interpose, at the time $\delta \tau$=4.8$\cdot \tau_S$, a detector ACROSS the kaons trajectory in order to measure whether a kaon is present or not by then. In the case that one is sure that a kaon was emitted at time 0, this measurement would be quasi-instantaneous in time, and would provide a {\it ``kaonic''} counterpart of the nearly-instantaneous photon-polarisation filtering described before.

    As was outlined in ref.\cite{bramon}, whenever the experimenter chooses to interpose ACROSS the way of the kaon a slab of matter (which constitutes a $\mathrm{K}^{0}$ preparation or filtering), or a kaon detector at a time $\delta \tau$=4.8$\cdot \tau_S$ in order to perform a kind of negative result measurement in the CP-eigenbasis, the measurement can be said to be {\it active} because it requires the free will of the experimentalist.
    
    It is worth noting that if one would measure the presence of a kaon at time 0, when a superposition of short-lived and long-lived states is created, we are certain to observe a kaon, but this doesn't tell us anything about its CP eigenvalue. On the contrary, a filtering in polarisation can be realized at any place along the propagation axis of light. Considered so, the analogy between quasi-spin and polarisation is potentially misleading and we ought to be careful when we claim that a quasi-spin measurement is performed at time $t$ (we come back to this point in appendix 3). This illustrates the main difficulty that is met in high energy physics, which is the lack of available experimental resources necessary for choosing at will the preparation and/or measurement bases.    
    In the practice, measurements are most often passive in the sense that detectors are put along the way of the kaons in order to collect particles that are spontaneously emitted during their decay, without having any influence on the time at which the decay occurs.

   {\bf Remark.} Following the discussion of the previous paragraph, it is clear that none of the measurements that are performed in a lab. (active or passive) can be considered to be instantaneous.
    Now, the basic ingredient of the so-called Quantum Zeno paradox \cite{misrasud,chiuchiu} is to admit that infinitely short measurements are possible. We show in appendix 2 that, not surprisingly, the Zeno paradox is neutralised when we formulate instantaneous measurements as continuous but nearly instantaneous measurements along the lines of section \ref{standard}. 
Let us now focus on the Temporal Wave Function approach, still in the absence of CP violation.

\subsection{Temporal Wave Function approach (Extended in Space Spin 1/2 Wave Function analogy).\label{wavefuc}} Let us consider \cite{09temporal} that  at time $T$ we prepare a particle in a coherent superposition state of a spin-up state $(1,0)$ localised over the positive semi-axis $0\leq x\leq+\infty$ according to a monochromatically modulated exponential distribution of ''Small'' extent, and of a spin-down state $(1,0)$ localised over the same region according to a monochromatically modulated exponential distribution of ''Large'' extent. The two-components of the Pauli wave function $(\Psi_1(x,T),\Psi_2(x,T))$ associated to this spinor are then equal to

\begin{equation}\Psi_1(x,T)=\psi_1(T)\cdot \sqrt{\lambda_S} e^{-\mathrm{i}(k_S-\frac{\mathrm{i}}{2}\lambda_S) x} \end{equation}

\begin{equation}\Psi_2(x,T)=\psi_2(T)\cdot \sqrt{\lambda_L}e^{-\mathrm{i}(k_L-\frac{\mathrm{i}}{2}\lambda_L) x},\end{equation}

where $\psi_1(T)$ and $\psi_2(t)$ are normalized complex amplitudes ($|\psi_1(T)|^2+|\psi_2(T)|^2=1$).

The probability to find spin ''up'' ((1,0)) at time $T$ in the interval $[x,x+dx]$ is, according to the usual rules of Quantum Mechanics, equal to $dx$ times $|\psi_1|^2\cdot \lambda_Se^{-\lambda_Sx}$
 Similarly, the probability to find spin ''down'' ((1,0)) at time $T$ in the interval $[x,x+dx]$ is equal to $dx$ times $|\psi_2|^2\cdot \lambda_Le^{-\lambda_Lx}$.
 
 In accordance with the Temporal Wave Function (TWF-O) approach \cite{09temporal}, let us now substitute space ($x$) by time ($t$), $\psi_1(x,T)$ by $\psi_1(t)$, $\psi_2(x,T)$ by $\psi_2(t)$ following the substitution rules (i,ii,iii) discussed in the introduction,and let us also substitute $\lambda_{S(L)}$ by $\Gamma_{S(L)}$ and $k_{S(L)}$ by $\omega_{S(L)}$, we obtain then a Temporal Wave Function model for the decay process that satisfies equations (\ref{fh2},\ref{fh3},\ref{fh4}). In particular, Eqns.(\ref{exprategen},\ref{exprategen2}), are satisfied in both approaches.
 
 This is not astonishing because  in absence of CP violation decay rates are exponential componentwise, a situation for which expressions (\ref{exp}) and (\ref{exprategen}) are equally valid.

Thus, at the level of approximation considered here it does not make any difference to adopt the traditional, standard, quasi-spin picture or the Temporal (spinorial) Wave Function (TWF-O) picture.

It is worth noting that in the case of CP-violation the situation is more complex because then we may no longer neglect the off-diagonal terms in the mass-decay matrix so that interferences appear and the decay is no longer purely exponential. Correspondingly, the expressions (\ref{exp}) and (\ref{exprategen}) are no longer compatible. Actually they are compatible only if the decay is exponential\footnote{In the case of bipartite systems the situation is more complex and as we shall see it can occur that both expressions are compatible also when non-exponential decay is present.} because if both expressions (\ref{exp}) and (\ref{exprategen}) were simultaneously satisfied one would get, taking the ratios of both equations,
\begin{equation}{P_s(t)\over p_d(t)}={P_s(t)\over-dP_s(t)/dt}={\rm a\ constant\ factor}\label{contr}.\end{equation}

Integrating (\ref{contr}) we would find that $P_s$ decreases exponentially with time. So, it is only for purely exponential decays that it does not make any difference (up to a constant in time calibration factor) to choose to express the decay rate through expressions (\ref{exp}) and (\ref{exp2}) or through (\ref{exprategen}) and  (\ref{exprategen2}). In appendix 1 we show by other methods that in the case of purely exponential decay processes the survival probability and decay rate approaches coincide. Now, we don't agree with the conclusion of reference \cite{girad} where it is said that all decay processes are, For All Practical Purposes, exponential. There exist many situations in which the decay is not exponential, for instance in presence of CP-violation as we shall discuss now.

\section{Two models for CP violation.}
\subsection{Preliminary discussion\label{stand}: Integrated versus Instantaneous Decay Rate.}
Even in the standard approach, some ambiguity remains present regarding what is exactly being measured during experimental tests of CP-violation.

It is clear that if the timing of arrival of the kaons was perfectly calibrated and controlled and also that the detector itself only registered decay products that are produced in a very narrow window of time, what is measured is the Decay Rate $p_d$ and not the Survival probability $P^s$ (equal to minus the Integrated Decay Rate).

Now, no detector is infinitely thin and what is measured is always an integrated decay rate, integrated over a time $\tau$ of the order of the typical dimension $L$ of the detector divided by $v_{kaons}$ the kaon velocity. Provided this time $\tau$ is small in comparison to the decay times $\tau_S$ and/or $\tau_L$), the decay rate, integrated over $\tau$, is obviously proportional to the instantaneous decay rate\footnote{Even in the case that the velocity of the kaons is close to $c$ this is still true because in the lab-reference frame lifetimes get {\bf dilated} by a relativistic $\gamma$ factor \cite{griffith}.}.

From our study of the literature and our discussions with specialists of CP-phenomenology, it is however not clear how to interpret the experimental tests of CP-violation\footnote{Typically, such tests occur as follows. A series of detections is
performed at various distances from the source of a neutral kaons (or meson pairs) in order to estimate the variation of the populations of decay products, for instance of emitted pion $\pi^+,\pi^-$ pairs in function of the proper time \cite{christ,frascati}. }. We are thus not able presently to answer with absolute certainty to questions such as

{\it What has been measured?... 
Is it an integrated decay rate and in this case what is the value of the integration time $\tau$? 
...Or is it the instantaneous decay rate?}

The reason therefore is that a measurement is never ideal and that it always requires some deconvolution procedure. In the case of CP-violation, a lot of noise is present and the deconvolution process as well as the data treatment are far from being simple and transparent. Moreover, the foundational aspects of the measurement process are usually skipped in standard textbooks on the subject and it is not easy to guess whether the authors consider that the experimentally measured quantity is the decay rate or the survival probability (see also section \ref{xxx}). We scrutinized the interpretation that {\it theoreticians} gave in the past of the standard formalism, and we found that it is commonly accepted to rely on the interpretation according to which the modulus squared of the effective wave function represents the Survival Probability (minus the Integrated Decay Rate) \cite{girad,bertlmann}. On the contrary, experimentators and phenomenologists often rely implicitly on an {\it ``hybrid''} view; we mean hereby that in several papers \cite{frascati,PRA63} and books \cite{perkins} it is neither the alternative expression (\ref{exprategen}) that is used nor the standard formula (\ref{exp}) but ``something in between'' that we shall illustrate by an explicit example soon (we also discuss this point in depth in the third appendix). 

In order to illustrate the fundamental incompatibility between all these approaches let us consider a representative example, the coherent superposition of two exponential decay processes, that was treated in the Time Superoperator formalism in ref.\cite{09superop}.

\begin{equation}\label{superplus}
\psi(t)=\left( \alpha_1e^{-\mathrm{i} (m_1-\frac{\mathrm{i}}{2}\Gamma_1)
t}+ \alpha_2 e^{-\mathrm{i}
(m_2-\frac{\mathrm{i}}{2}\Gamma_2) t}\right),
\end{equation}
where the $\alpha$ factors are complex amplitudes assigned to each of the decay processes: $\alpha_i=|\alpha_i|e^{-i\phi_i}$, $i=1,2$, $|\alpha_1|^2+|\alpha_2|^2=1.$
The modulus square of $\psi$ is equal to

\begin{equation}\label{super}|\psi(t)|^2=| \alpha_1|^2e^{-\Gamma_1t}+2| \alpha_1||\alpha_2|e^{(-{\Gamma_1+\Gamma_2\over 2}t)}cos(\triangle mt+\triangle \phi)+|\alpha_2|^2e^{-\Gamma_2t},\end{equation} where $\triangle \phi=\phi_1-\phi_2$.
In the standard approach, $|\psi(t)|^2$ represents $P_s(t)$ the survival probability at time $t$. 

Then the decay rate obeys

\begin{eqnarray}p^{standard}_d(t)={-d|\psi(t)|^2\over dt}=| \alpha_1|^2\Gamma_1e^{-\Gamma_1t}+|\alpha_2|^2\Gamma_2e^{-\Gamma_2t}\nonumber\\+2| \alpha_1||\alpha_2|e^{(-{\Gamma_1+\Gamma_2\over 2}t)}({\Gamma_1+\Gamma_2\over 2}cos(\triangle mt+\triangle \phi)+\triangle m sin(\triangle mt+\triangle \phi)
)\nonumber\\=| \alpha_1|^2\Gamma_1e^{-\Gamma_1t}+|\alpha_2|^2\Gamma_2e^{-\Gamma_2t}+2| \alpha_1||\alpha_2|e^{(-{\Gamma_1+\Gamma_2\over 2}t)}Rcos(\triangle mt+\triangle \phi+\psi),\label{superstandard}\end{eqnarray} 
where $R$ and $\psi$ are real and $Re^{i\psi}={\Gamma_1+\Gamma_2\over 2}-i\triangle m$.

In the {\it ``hybrid''} approach, it is implicitly assumed that the corresponding decay rate, the pdf $p_d(t)$ is proportional to $|\psi(t)|^2$ so that
\begin{equation}\label{hybrid}p^{hybrid}_d(t)=cst.|\psi(t)|^2=cst.(| \alpha_1|^2e^{-\Gamma_1t}+2| \alpha_1||\alpha_2|e^{(-{\Gamma_1+\Gamma_2\over 2}t)}cos(\triangle mt+\triangle \phi)+|\alpha_2|^2e^{-\Gamma_2t}),\end{equation}We think that this view is inconsistent, because for instance in the case that decoherence is present so that the phase difference $\triangle \phi$ is a random quantity, the interference modulation disappears and we find in the standard approach that the survival probability is the weighted sum 
$|\alpha_1|^2e^{-\Gamma_1t}+| \alpha_2|^2e^{-\Gamma_2t}$ which is a natural result. On the contrary, if it is assumed as in the hybrid approach that $|\psi(t)|^2$ is proportional to the decay rate, then the survival probability obeys, in presence of full decoherence, $P_s(t)={|\alpha_1|^2\over \Gamma_1}e^{-\Gamma_1t}+{|\alpha_2|^2\over \Gamma_2}e^{-\Gamma_2t}$ so that the weight of the shortest process is boosted in a ratio $\tau_S/\tau_L$, which is, in our view, an artifical and counterintuitive result (see also our discussion in the second part of appendix 3).

In the TWF-O approach on the contrary, the decay rate obeys an expression of the form

\begin{equation}p^{Time Operator}_d(t)=| \alpha_1|^2\Gamma_1e^{-\Gamma_1t}+2| \alpha_1||\alpha_2|\sqrt{\Gamma_1\Gamma_2}e^{(-{\Gamma_1+\Gamma_2\over 2}t)}cos(\triangle mt+\triangle \phi)+|\alpha_2|^2\Gamma_2e^{-\Gamma_2t},\label{timeoperator}\end{equation} 
as we shall show soon, so that in the case of full decoherence it leads to the same predictions as the standard approach.

The main message of our paper is that {\bf the expressions (\ref{superstandard}), (\ref{hybrid}) and 
(\ref{timeoperator}) are incompatible}. This is because, although they all predict a superposition of two purely exponential processes and of an exponentially decreasing periodic interference term, with the same lifetimes and periods, they differ in the respective weights assigned to these respective contributions\footnote{Actually the main motivation that underlies the present analysis comes from our study of the SuperTimeOperator \cite{09superop} where we noted also a discrepancy of the weight of the ``Long'' exponential process regarding ``hybrid'' predictions.}. Therefore the different interpretations can, in principle, be discriminated experimentally.

This kind of questions could be addressed in future experiments about CP-violation as we shall show in the next section but they are also likely to be tested with atomic and/or quantum optical metastable states, for which the freedom to choose at will the preparation and mesurement bases as well as the degree of precision and of control are quite higher than in high energy particles physics. In particular during the last ten years, the perspective of a quantum computer and the quest of always better metrology standards stimulated an impressive technological development (at the level of ion traps \cite{iontraps}, QED cavities \cite{haroche}, spectroscopy \cite{wineland,hansche} and so on).

One could object that in the domain of high energy physics plenty of experiments were already performed in the past showing no discrepancy between standard predictions and experiments (also in the case of meson oscillations, EPR mesonic states and so on) but the situation is not so simple.

Actually in this type of experiments a lot of background noise is present and it is not so easy to deconvolute the real physical signal from the response of the detectors and from various parasite processes. For instance, the measure of the survival probability makes it easy to measure the phase of the violation parameter because oscillations surimposed on a noisy background are rather easy to isolate and one can measure a phase with a rather good precision, but the amplitude of these oscillations is difficult to estimate because it is difficult to estimate the relative calibration of the detectors and the background noise contribution. This difficulty is due among others to the fact that the experimental set up is so heavy that in general sources as well as detectors are unique, which prohibits mutual calibration procedures that require several different detectors.

In the next section we shall reconsider the Cronin and Fitch experiment \cite{christ}, having in mind the discussion above.


\subsection{Cronin and Fitch experiment: Standard approach.\label{surv}}
In the celebrated Cronin and Fitch experiment, the initial state is a neutral kaon state 

$|\mathrm{K}^0\rangle$=$ \frac{1}{\sqrt{2}}\big{(}|\mathrm{K}_1\rangle
+|{\mathrm{K}}_2\rangle\big{)}$=$ \frac{\sqrt{1+|\epsilon|^2}}{\sqrt{2}(1+\epsilon)}\big{(}|\mathrm{K}_S\rangle
+|{\mathrm{K}}_L\rangle\big{)}$.

 Then, the number of pairs and triplets of pions that are produced during the decay of the kaon is measured, at a certain distance from the source. As the $\pi^+,\pi^-$ pairs are CP-eigenstates 
for the eigenvalue +1, their presence reveals that the $\mathrm{K}_1$ state is ``populated''. In the case that $\epsilon$ equals 0 (no off-diagonal terms in the mass-decay matrix and no CP-violation), $|\mathrm{K}_S\rangle=|\mathrm{K}_1\rangle$ and $|\mathrm{K}_L\rangle=|\mathrm{K}_2\rangle$ so that, for times quite longer than $\tau_S$, no $\pi^+,\pi^-$ pair is likely to be observed, in accordance with the model of continuous in time filtering that was discussed in the section \ref{contmeas}. The experiment
shows on the contrary that these pairs are observed which is a manifestation of CP-violation. 

The standard modelisation of the process goes as follows: in accordance with the expression (\ref{kk2}) and in virtue of the linearity of Schr\"odinger equation, we find that at time $t$ the state obeys 

\begin{eqnarray}|\Psi(t)\rangle=\psi_1(t)|\mathrm{K}_1\rangle+\psi_2(t)|\mathrm{K}_2\rangle\nonumber\\
=\frac{1}{\sqrt{2}(1+\epsilon)} \left(  \left( e^{-\mathrm{i} (m_S-\frac{\mathrm{i}}{2}\Gamma_S)
t}+ \epsilon e^{-\mathrm{i}
(m_L-\frac{\mathrm{i}}{2}\Gamma_L) t}\right)|\mathrm{K}_1\rangle
+\left( \epsilon e^{-\mathrm{i} (m_S-\frac{\mathrm{i}}{2}\Gamma_S)
t}+  e^{-\mathrm{i}
(m_L-\frac{\mathrm{i}}{2}\Gamma_L) t}\right)
     |\mathrm{K}_2\rangle \right)\label{scramble}\end{eqnarray}

The amplitude $\psi_1(t)$ of the projection onto the CP=+1 sector (so to say onto the $\mathrm{K}_1$ basis-state) at time $t$ obeys thus
\begin{eqnarray}\psi_1(t)=\frac{\sqrt{1+|\epsilon|^2}}{\sqrt{2}(1+\epsilon)}(\langle K^1|\mathrm{K}_S\rangle e^{-\mathrm{i}E_St}+ \langle K^1|\mathrm{K}_L\rangle e^{-\mathrm{i}E_Lt})\nonumber\\\label{psi}
=\frac{1}{\sqrt{2}(1+\epsilon)}\left( e^{-\mathrm{i} (m_S-\frac{\mathrm{i}}{2}\Gamma_S)
t}+ \epsilon e^{-\mathrm{i}
(m_L-\frac{\mathrm{i}}{2}\Gamma_L) t}\right)
\end{eqnarray} 

According to the standard approach, it is the expression  (\ref{exp}) (considered componentwise) that describes the {\it integrated} decay rate and we get by a direct computation, projecting onto the CP=+1 sector\footnote{We project onto the CP=+1 sector because we only consider here the measurements of pairs of pions which are CP=+1 states.},
\begin{eqnarray}P^1_s(t)&=&P^1_s(0)\cdot{\psi_1^*(t)\psi_1(t)\over \psi_1^*(0)\psi_1(0)}\nonumber\\
 &=&{P^1_s(0)\over |1+\epsilon|^2}\,\bigg{(}e^{-\Gamma_S t}+
|\epsilon |^2e^{-\Gamma_L t}+
2|\epsilon |e^{-({\Gamma_S +\Gamma_L\over 2})t} \cos
\big{(}\triangle m t+\arg(\epsilon)\big{)}\bigg{)},
\label{intens}\end{eqnarray}
 
 
where $\triangle m=|m_L-m_S|$.   According to the expression (\ref{intens}), two effects are likely to occur that reveal  the existence of a CP violation. For times intermediate between $\tau_S$ and $\tau_L$ an interference term is likely to appear. Besides, for times of the order of and larger than $\tau_L$ decay is still likely to occur in the CP=+1 sector with probability $|\epsilon |^2$. This last effect is what was observed in the Cronin and Fitch experiment  \cite{christ}. By fitting this oscillating contribution with the observed data one
derives an estimation of the mass difference between the short and
long lived state as well as the phase of $\epsilon$
and its amplitude. All this leads to an experimental estimation of
$\epsilon$ (that we shall denote $\epsilon^{\mathrm{exp}}$) \cite{perkins}
\begin{equation}\label{expepsilon}
|\epsilon^{\mathrm{exp}}|=(2.27\pm0.02)\times10^{-3}, ~~~
\mathrm{arg}(\epsilon^{\mathrm{exp}})=43.37^\circ.
\end{equation}


Before we study the predictions made in the framework of the TWF-O approach concerning CP violation it is useful to derive the {\it instantaneous} decay rate (pdf) from the expression (\ref{intens}) which gives

\begin{eqnarray}&p^1_d(t)&=-{dP^1_s(t)\over dt}=P^1_s(0)\cdot{\psi_1^*(t)\psi_1(t)\over \psi_1^*(0)\psi_1(0)}={\Gamma_S P^1_s(0)\over |1+\epsilon|^2}\,\bigg{(}e^{-\Gamma_S t}+
|\epsilon |^2{\Gamma_L \over \Gamma_S} e^{-\Gamma_L t}\nonumber\\
&+&2{|\epsilon |\over \Gamma_S}({\Gamma_S +\Gamma_L\over 2}e^{-({\Gamma_S +\Gamma_L\over 2})t} \cos
\big{(}\triangle m t+\arg(\epsilon)\big{)}+\triangle me^{-({\Gamma_S +\Gamma_L\over 2})t} \sin
\big{(}\triangle m t+\arg(\epsilon)\big{)}\bigg{)}\nonumber\\
&\approx&  {\Gamma_S P^1_s(0)\over |1+\epsilon|^2}\,\bigg{(}e^{-\Gamma_S t}+
|\epsilon |^2{\Gamma_L \over \Gamma_S} e^{-\Gamma_L t}+
2{|\epsilon |\triangle m\over \Gamma_S}(e^{-({\Gamma_S +\Gamma_L\over 2})t} \cos
\big{(}\triangle m t+\arg(\epsilon)\big{)}\nonumber\\ &+&e^{-({\Gamma_S +\Gamma_L\over 2})t} \sin
\big{(}\triangle m t+\arg(\epsilon)\big{)}\bigg{)}\nonumber\\
&\approx&  {\Gamma_S P^1_s(0)\over |1+\epsilon|^2}\,\bigg{(}e^{-\Gamma_S t}+
|\epsilon |^2{\Gamma_L \over \Gamma_S} e^{-\Gamma_L t}+
{|\epsilon |\over \sqrt 2}e^{-({\Gamma_S +\Gamma_L\over 2})t} \cos
\big{(}\triangle m t+\arg(\epsilon)-{\pi\over 4}\big{)}\bigg{)}\label{intensementbon}
\end{eqnarray}

where we made use of the fact that in the case of kaons, $\triangle m\approx {\Gamma_S+\Gamma_L\over 2}\approx {\Gamma_S\over 2}$.

\subsection{Temporal Wave Function approach (Extended in Space Spin 1/2 Wave Function analogy).\label{papa}} 

Let us assume (in accordance with the results presented in reference \cite{09temporal}) that at time $T$ we prepare a particle in a fifty-fifty coherent superposition of 

1) a spin state ${1\over 1+|\epsilon|^2}(1,\epsilon)$ localised over the positive semi-axis $0\leq x\leq+\infty$ according to a monochromatically modulated exponential distribution of ''Small'' extent, and of 

2) a spin state ${1\over 1+|\epsilon|^2}(\epsilon,1)$ localised over the same region according to a monochromatically modulated exponential distribution of ''Large'' extent.

 The two-components of the Pauli wave function $(\Psi_1(x,T),\Psi_2(x,T))$ associated to this spinor being equal to

\begin{equation}\Psi_1(x,T)={1\over \sqrt{2}\tilde{N}}\cdot (\sqrt{\lambda_S}e^{-\mathrm{i}(k_S-\frac{\mathrm{i}}{2}\lambda_S) x} +\epsilon \sqrt{\lambda_L}
e^{-\mathrm{i}(k_L-\frac{\mathrm{i}}{2}\lambda_L) x})\label{kwak}\end{equation}

\begin{equation}\Psi_2(x,T)={1\over \sqrt{2}\tilde{N}}\cdot( \epsilon \sqrt{\lambda_S}e^{-\mathrm{i}(k_S-\frac{\mathrm{i}}{2}\lambda_S) x}
+\sqrt{\lambda_L}e^{-\mathrm{i}(k_L-\frac{\mathrm{i}}{2}\lambda_L) x})\label{kwak2}\end{equation}

The probability to find spin ''up'' ((1,0)) at time $T$ in the interval $[x,x+dx]$ is, according to the usual rules of Quantum Mechanics, equal to ${dx\over 2\tilde{N}^2}$ times

 \begin{equation}(\lambda_Se^{-\lambda_Sx}+|\epsilon|^2\lambda_L e^{-\lambda_Lx}+2Re(\epsilon \sqrt{\lambda_S\cdot \lambda_L}e^{-(i(k_L-k_S)+{\lambda_S+\lambda_L\over 2})x})).\end{equation}
 Similarly, the probability to find spin ''down'' ((0,1)) at time $T$ in the interval $[x,x+dx]$ is equal to ${dx\over 2\tilde{N}^2}$ times

 \begin{equation}(|\epsilon|^2\lambda_Se^{-\lambda_Sx}+\lambda_Le^{-\lambda_Lx}+2Re(\epsilon \sqrt{\lambda_S\cdot \lambda_L}e^{-(i(k_S-k_L)+{\lambda_S+\lambda_L\over 2})x})),\end{equation}
 with $\tilde N$ chosen in order to normalise the probability of presence to 1 which imposes that
 
 $\tilde{N}^2=1+|\epsilon|^2+Re(\epsilon){(\sqrt{\lambda_S\cdot \lambda_L})(\lambda_S+\lambda_L)\over (\delta k)^2+({\lambda_S+\lambda_L\over 2})^2}$.
 
 According to the approach of section \ref{wavefuc}, this model can serve us to simulate CP-violation, as we have shown in ref.\cite{09temporal}, provided we substitute in the expressions (\ref{kwak},\ref{kwak2}) the parameters $x$, $k$ and $\lambda$ by $v\cdot t$, $m/v$ and $\Gamma/v$ respectively (in our system of units $h$ and $c$ are taken to be equal to unity).

Making then use of the expressions (\ref{exprategen},\ref{kwak}) the theoretically estimated production rate of pion pairs $I(t)$ has now the following form:
\begin{eqnarray}\label{in2}
 I(t) ={I_0\over |1+\epsilon\,\,\sqrt{\frac{\Gamma_L}{\Gamma_S}} |^2} \bigg( \,e^{-\Gamma_S t}+
|\epsilon\,\,\sqrt{\frac{\Gamma_L}{\Gamma_S}}|^2\, e^{-\Gamma_Lt}+ 2
|\epsilon\,\,\sqrt{\frac{\Gamma_L}{\Gamma_S}}|\, e^{-{
\Gamma_S+\Gamma_L\over 2}t}\cos(\triangle m
t+\arg(\epsilon))\bigg).
\end{eqnarray}

In the next sections we shall show that the Temporal Wave Function approach leads to new, non-standard, predictions that could be tested experimentally, in the single particle case (section \ref{single}) and in the EPR pair case (section \ref{cici}).

\section{Discrepancies between the standard approaches and the Temporal Wave Function approach in presence of CP 
violation.\label{single}}
\subsection{Experimental proposal.}
On the basis of the results of the last section one can conceive a crucial experiment that would allow us to discriminate the Temporal Wave Function (or Time Operator) approach and the standard approach. As we shall now show, if we reproduce the conditions of the Cronin and Fitch experiment, we are able in principle to discriminate both approaches.

In order to do so, let us compare the standard and non-standard (TWF-O) predictions concerning the rate of production of pion pairs\footnote{In a previous paper \cite{09temporal} we compared the temporal wave function and the ``hybrid'' approach and noted that formally the expressions (\ref{hybrid}) and (\ref{timeoperator}) are indistinguishible provided the complex amplitudes $\alpha_1$ and $\alpha_2$ are ``renormalised'' in an ad hoc manner. In the case considered here (the Cronin and Fitch experiment) this leads to the prescription $\epsilon^{\mathrm{ren.}}=\epsilon^{\mathrm{exp.}}$ so to say $\epsilon=\epsilon^{\mathrm{exp.}}\cdot\sqrt{{\Gamma_S\over \Gamma_L}}, \label{renormalis}$ which is, in the case of neutral kaons, more or less 30 times larger than the usual prediction $\epsilon=\epsilon^{\mathrm{exp.}}\approx 2.27\,10^{-3}$. Actually we also estimated the value of $\epsilon$ making use of the Friedrich's model in that publication \cite{09temporal} but this kind of refinement is clearly out of the scope of the present paper.}.

The standard prediction for the rate of detection of pion pairs is 
\begin{eqnarray}I(t)=cst.p^1_d(t)
\approx  {I(0)\over |1+\epsilon|^2}\,\bigg{(}e^{-\Gamma_S t}+
|\epsilon |^2{\Gamma_L \over \Gamma_S} e^{-\Gamma_L t}+
{|\epsilon |\over \sqrt 2}e^{-({\Gamma_S +\Gamma_L\over 2})t} \cos
\big{(}\triangle m t+\arg(\epsilon)-{\pi\over 4}\big{)}\bigg{)}
\label{intensementbon2}\end{eqnarray}

in virtue of equation (\ref{intensementbon}), which is a special case of the more general expression (\ref{superstandard}).

Obviously the expression (\ref{intensementbon2}) is incompatible with the corresponding expression (\ref{in2}) that we derived in the temporal wave function or time operator approach, which is a special case of the more general expression (\ref{timeoperator}).

This is because there exists no choice for $\epsilon$ that will allow us to obtain a perfect fit between (\ref{intensementbon2}) and (\ref{in2}). Indeed, the ratio between the square root of the weight of the factor $e^{-\Gamma_L t}$ and the weight of the oscillating interference term differs in the expressions (\ref{intensementbon2}) and (\ref{in2}).

This ratio is a measurable quantity so that its precise experimental estimation should in principle allow us to discriminate between the standard and temporal wave-function approaches...

One could object that the experiment has been realized already in 1964 but the situation is not so simple as we shall discuss now.

\subsection{What has been measured in the Cronin and Fitch experiment?\label{xxx}}

In many standard text-books (like the Perkins \cite{perkins} for instance) the production rate of pion pairs  observed in the Cronin and Fitch experiment is assumed to obey 

  \begin{eqnarray}\label{perkins}
\nonumber I(t)={I_0\over |1+\epsilon|^2}\,\bigg{(}e^{-\Gamma_S t}+
|\epsilon |^2e^{-\Gamma_L t}+
2|\epsilon |e^{-({\Gamma_S +\Gamma_L\over 2})t} \cos
\big{(}\triangle m t+\arg(\epsilon)\big{)}\bigg{)}
\end{eqnarray}
where $I_0$ is a global factor constant in time.

It is easy to check that this expression perfectly fits with the expression (\ref{hybrid}) so that it can be characterized as an ``hybrid'' expression, incompatible with both the standard approach and the temporal wave function approach as we discussed in the section \ref{stand}.


Besides, we scrutinized the original paper of 1964 by Christenson {\it et al.} \cite{christ} in order to understand what has been measured and how the results were interpreted by then. Essentially the Cronin and Fitch experiment consisted of measuring the ratio $R$ between the number of charged pion pairs ($\pi^+, \pi^-$) and triplets at a (proper) time quite longer that $\tau_S$ (this time was so long that \cite{griffith} $e^{-\Gamma_St}/e^{-\Gamma_Lt}\approx ({1\over 500})^{17}\approx 1,3\cdot 10^{-46}\approx 0$).

During the experiment 45 pairs were observed and 22700 decays occured so that 

$R={45\over 22700}=(2.0\pm 0.4)10^{-3}$. 

According to the authors, the relation between this ratio and $|\epsilon|$ obeys\begin{equation}|\epsilon|^2=R_T {\tau_1\over \tau_2},\label{fichtre}\end{equation} where $R_T={3\over 2} R$ and ${\tau_1\over \tau_2}$ is equal (in the notation of 1964) to the ratio between the Short and Long lifetimes. The correction factor ${3\over 2}$ is explained to be due to the fact that decay in the CP=+1 sector branches to charged pion pairs with probability 2/3 (the remaining 1/3 corresponding to neutral pion pairs that were not detected). Substituting the measured value of $R=2\cdot 10^{-3}$ into the expression (\ref{fichtre}) we get

$|\epsilon|^2={3\over 2}R{\tau_1\over \tau_2}={3\over 2}\cdot2.0 \cdot10^{-3}\cdot {8.92\cdot 10^{-11}\over 5.17\cdot 10^{-8}}$ so that finally we find\footnote{By the way, we corrected {\it en passant} a typo present in ref.\cite{christ} that could maybe explain partially the confusion that exists in the literature regarding the subject: the authors wrote $|\epsilon|^2=R_T \tau_1 \tau_2$ which is inconsistent dimensionally and above all doesn't lead to the right estimation of $|\epsilon |$.}  $|\epsilon|^2=5.2\cdot 10^{-6}$ which corresponds to the CP violation $\epsilon=2.3\cdot 10^{-3}$ reported in ref.\cite{christ}, in agreement with the commonly accepted value mentioned in equation (\ref{expepsilon}). 

Actually, the reason of the choice of the corrective factor ${\tau_1\over \tau_2}$=$({\Gamma_S\over \Gamma_L})^{-1}$ in ref.\cite{christ} is not clear to us: obviously the experiment is such that what is measured is a ratio between two decay rates, the one associated to the pair production and the one associated to the triplet production. We are indeed in right to talk about decay rates here because the decay products were collected over a time $\tau$ obviously very small in comparison to the decay time $\tau_L$.

Let us now reconsider the experiment at the light of our previous analysis. Projecting the state at time $t$ onto the CP=+1 and -1 sectors and neglecting the factors that contain contributions in $\Gamma_S t$ we get

\begin{equation}\label{psi1}
\psi_1(t)=\frac{1}{\sqrt{2}(1+\epsilon)}\left(  \epsilon e^{-\mathrm{i}
(m_L-\frac{\mathrm{i}}{2}\Gamma_L) t}\right)
\end{equation}
\begin{equation}\label{psi2}
\psi_2(t)=\frac{1}{\sqrt{2}(1+\epsilon)}\left( e^{-\mathrm{i}
(m_L-\frac{\mathrm{i}}{2}\Gamma_L) t}\right)
\end{equation}

Formulated in terms of the expression (\ref{superplus}) this would mean that we neglect the contributions where $\alpha_1$ appears and substitute ($\alpha_2, \Gamma_2$) by ($\alpha_L, \Gamma_L$), which yields depending our choice to follow the standard (\ref{superstandard}) hybrid (\ref{hybrid}) or TWF-O (\ref{timeoperator}) approaches to the expressions

$p_{d}^{standard}(pairs)\approx |\epsilon|^2\Gamma_L$, $p_{d}^{hybrid}(pairs)\approx |\epsilon|^2$ and $p_{d}^{TimeOperator}(pairs)\approx |\epsilon|^2\Gamma_L$ while 

 $p_{d}^{standard}(triplets)\approx \Gamma_L$, $p_{d}^{hybrid}(triplets)\approx 1$ and $p_{d}^{TimeOperator}(triplets)\approx \Gamma_L$.

Henceforth, the three approaches agree concerning the prediction of $|\epsilon|^2$:

$R_T^{standard}=R_T^{hybrid}=R_T^{Time Operator}=|\epsilon|^2$.

Unfortunately a factor ${\tau_1\over \tau_2}$ is missing if we compare to the expression (\ref{fichtre}).

It could be related to the calibration mentioned in ref.\cite{christ} that was performed with tungsten in order to estimate the detection response of pion pairs but, anyhow, the appearance of this factor ${\tau_1\over \tau_2}$=$({\Gamma_S\over \Gamma_L})^{-1}$ remains mysterious in our eyes. In appendix 3 we propose a plausible way to motivate the appearance of this mysterious factor\footnote{It is interesting to note that if we neglect this factor, the amplitude of $\epsilon$ gets multiplied by a factor of the order of 30, in good agreement with predictions that we obtained in the framework of the Friedrichs model \cite{fried} in the past \cite{CDS,cds2}.}.


\section{Comparison between the standard approach and the temporal wave-function approach in the case of EPR-Bohm meson pairs.\label{cici}}
\subsection{Singlet state.}

EPR correlations exhibited by entangled mesons  prepared in the ``singlet'' Bell state \cite{bellstate} have been measured experimentally in several labs, (for instance in Frascati \cite{frascati} and in Geneva (CP-Lear) \cite{cplear} with kaons ($K$), and in Tsukuba (KEK) \cite{tsukuba} with $B$ mesons) and therefore it would be interesting to estimate the predictions of our wave function model in order to compare them with the standard predictions.

 In this section we show that no discrepancy exists between predictions made in the standard approaches and those made in the temporal wave function approach regarding EPR correlations: the two approaches coincide in the case that the entangled state is a so-called {\it singlet} state. Moreover, there exists an infinity of entangled states for which this is also true as we show in the next section (\ref{nosinglet}).

 In the $\phi$ factory at Frascati for instance, kaon pairs are produced in the  EPR-Bohm (so-called {\it singlet}) state
 
 \begin{eqnarray}|\psi\rangle={1\over \sqrt 2}(|{\mathrm{K}}^0\rangle_l|\overline{\mathrm{K}}^0\rangle_r-|\overline{\mathrm{K}}^0\rangle_l|\mathrm{K}^0\rangle_r)\label{mix}\nonumber\\
 ={1+|\epsilon|^2\over \sqrt 2(1-\epsilon^2)}\bigg{(}| \mathrm{K}_L\rangle_l \mathrm{K}_S\rangle_r - | \mathrm{K}_S\rangle_l | \mathrm{K}_L\rangle_r\bigg{)},\label{phi}\end{eqnarray}
where the indices $l$ and $r$ refer to the fact that those kaons are sent along opposite directions (with equal velocities $v$). Such a state possesses a total quasi-spin and a total momentum both equal to zero, due to conservation laws. Actually, entanglement and EPR correlations are seen here to be the consequence of conservation laws, a common situation in quantum physics.

The standard prediction for the survival probability of the pair of kaons at distances $d_l=v\cdot t_l$ on the left and $d_r=v\cdot t_r$ is, as a consequence of equation \ref{exp2}, and expanding the wave function in the CP product eigen-basis, equal to 

\begin{equation}P_S(t_l,t_r)=\|\psi(t_l,t_r)^2\|/\|\psi(0,0)^2\| \end{equation} with \begin{equation}\| \psi(t_l,t_r)\|^2=\sum_{i,j=1}^2\psi^*_{ij}(t_l,t_r)\psi_{ij}(t_l,t_r),\end{equation} where 

\begin{eqnarray}\psi_{ij}(t_l,t_r)=_l\langle\mathrm{K}_i| _r\langle\mathrm{K}_j|\psi(t_l,t_r)\rangle=\nonumber\\_l\langle\mathrm{K}_i| _r\langle\mathrm{K}_j|{1+|\epsilon|^2\over \sqrt 2(1-\epsilon^2)}(e^{-\mathrm{i} (m_L-\frac{\mathrm{i}}{2}\Gamma_L)
t_l}|{\mathrm{K}}_L\rangle_l e^{-\mathrm{i} (m_S-\frac{\mathrm{i}}{2}\Gamma_S)
t_r}|\mathrm{K}_S\rangle_r-e^{-\mathrm{i} (m_S-\frac{\mathrm{i}}{2}\Gamma_S)
t_l}|{\mathrm{K}}_S\rangle_l e^{-\mathrm{i} (m_L-\frac{\mathrm{i}}{2}\Gamma_L)
t_r}|\mathrm{K}_L\rangle_r)\nonumber\\
={1+|\epsilon|^2\over \sqrt 2(1-\epsilon^2)}(e^{-\mathrm{i} (m_L-\frac{\mathrm{i}}{2}\Gamma_L)
t_l}e^{-\mathrm{i} (m_S-\frac{\mathrm{i}}{2}\Gamma_S)
t_r} \langle\mathrm{K}_i|{\mathrm{K}}_L\rangle_l  \langle\mathrm{K}_j|\mathrm{K}_S\rangle_r
-e^{-\mathrm{i} (m_S-\frac{\mathrm{i}}{2}\Gamma_S)
t_l} e^{-\mathrm{i} (m_L-\frac{\mathrm{i}}{2}\Gamma_L)
t_r} \langle\mathrm{K}_i|{\mathrm{K}}_S\rangle_l  \langle\mathrm{K}_j|\mathrm{K}_L\rangle_r
\label{shishi}\end{eqnarray}

According to the distinction between active and passive measurements made in ref.(\cite{bramon}) (see also our discussion of section \ref{contmeas}), the Frascati experiment, in which the rates of production of pairs of pions are measured ALONG (and not ACROSS) the trajectories of the kaons in the left and right regions is a PASSIVE measurement.

It is natural to interpret the probability $P_S(t_l,t_r)$ as the sum of the probabilities that the pair survives without decaying in the CP=+1 or - 1 channels during the time interval $[0,t_l]$ in the left region and $[0,t_r]$ in the right region, while, accordingly, the probability that for instance the pair decays in the CP=+1 channel during the time intervals   $[t_l,t_l+\delta t_l]$ in the left  region and $[t_r,t_r+\delta t_r]$ in the right region is equal to $(\delta t_l\frac{\partial }{\partial t_l}+\delta t_r\frac{\partial }{\partial t_r})P^{11}_S(t_l,t_r)$, in the case that the temporal resolution $\delta t$ of the detector is small enough (sufficiently smaller than $\tau_S$ in this case).

  Assuming that detectors have the same dimensions at both sides, and also taking account of the fact that particles at both sides have same velocities so that $\tau_l=\tau_r$, this probability is also proportional (up to constant in time calibration and normalisation factors) to $p^{11}_d(t_l,t_r)=(\frac{\partial }{\partial t_l}+\frac{\partial }{\partial t_r})P^{11}_S(t_l,t_r)$. 
 A direct computation shows that 

\begin{eqnarray}p^{11}_d(t_l,t_r)=(\frac{\partial }{\partial t_l}+\frac{\partial }{\partial t_r})P^{11}_S(t_l,t_r)=\nonumber\\ (\frac{\partial }{\partial t_l}+\frac{\partial }{\partial t_r}){|\epsilon|^2\over  2(1-\epsilon^2)^2} 
|e^{-\mathrm{i} (m_L-\frac{\mathrm{i}}{2}\Gamma_L)
t_l}e^{-\mathrm{i} (m_S-\frac{\mathrm{i}}{2}\Gamma_S)
t_r} -
e^{-\mathrm{i} (m_S-\frac{\mathrm{i}}{2}\Gamma_S)
t_l} e^{-\mathrm{i} (m_L-\frac{\mathrm{i}}{2}\Gamma_L)
t_r}|^2
\nonumber\\ \approx (\Gamma_S+\Gamma_L)|\epsilon|^2  \cdot\bigg{(}
e^{-\Gamma_Lt_l-\Gamma_S t_r} +e^{-\Gamma_St_l-\Gamma_L t_r}
-2e^{-\frac{(\Gamma_S+\Gamma_L)(t_l+t_r)}{2}}
cos (\Delta m (t_l-t_r))\bigg{)}
\nonumber\\ =(\Gamma_S+\Gamma_L){|\epsilon|^2\over  2(1-\epsilon^2)^2} |e^{-\mathrm{i} (m_L-\frac{\mathrm{i}}{2}\Gamma_L)
t_l}e^{-\mathrm{i} (m_S-\frac{\mathrm{i}}{2}\Gamma_S)
t_r} -
e^{-\mathrm{i} (m_S-\frac{\mathrm{i}}{2}\Gamma_S)
t_l} e^{-\mathrm{i} (m_L-\frac{\mathrm{i}}{2}\Gamma_L)
t_r}|^2.\label{frascatipd}\end{eqnarray}

The former quantity is proportional (up to a calibration factor $|\langle \pi^+\pi^- | \mathrm{K}_1\rangle|^4  |$ aimed at converting the kaonic decay rate in the CP=+1 sector into the detection rate of pion pairs)
to what is called in ref.\cite{frascati} (expression (2)) ``{\it the decay intensity for the process}  $\phi\rightarrow ({\rm 2 neutral kaons})\rightarrow  \pi^+\pi^-$'', a property that can be generalised to a large class of entangled states as we show in the next section\footnote{\label{footnoteremark}Remarkably, in the case that the pair of Kaons is prepared in a {\it singlet} EPR-Bohm state of the type considered here, ${(\frac{\partial }{\partial t_l}+\frac{\partial }{\partial t_r})P^{11}_S(t_l,t_r)\over P^{11}_S(t_l,t_r)}$ is a constant factor that does not depend on time.}.

Let us now consider the prediction that we get in the temporal wave function (time operator) approach. 

In this approach, when at time 0 a $|\mathrm{K}_S\rangle$ is prepared, it is described by the quasi-spinorial temporal wave function
\begin{equation}\label{psitS}
(\psi^S_1(t),\psi^S_2(t))= {1\over 1+|\epsilon|^2}(1,\epsilon) \sqrt{\Gamma_S}e^{-\mathrm{i} (m_S-\frac{\mathrm{i}}{2}\Gamma_S)
t}
\end{equation}
Similarly a $\mathrm{K}_L$ state is associated to the bi-spinorial function ($\psi^L_1(t)$,$ \psi^L_2(t)$) through

\begin{equation}\label{psitL}
(\psi^L_1(t),\psi^L_2(t))= {1\over 1+|\epsilon|^2}(\epsilon,1) \sqrt{\Gamma_L}e^{-\mathrm{i} (m_L-\frac{\mathrm{i}}{2}\Gamma_L)
t}
\end{equation}
Combining (\ref{psitS},\ref{psitL}) with (\ref{phi}) we get

 \begin{eqnarray}|\psi\rangle(t_l,t_r)=
 {1+|\epsilon|^2\over \sqrt 2(1-\epsilon^2)}\bigg{(}\left(\begin{array}{c} 1\\ \epsilon
\end{array}\right)_l\otimes 
\left(\begin{array}{c} \epsilon \\ 1
\end{array}\right)_r \sqrt{\Gamma_S}e^{-\mathrm{i} (m_S-\frac{\mathrm{i}}{2}\Gamma_S)
t_l}\cdot  \sqrt{\Gamma_L}e^{-\mathrm{i} (m_L-\frac{\mathrm{i}}{2}\Gamma_L)
t_r}-\nonumber \\
\left(\begin{array}{c} \epsilon \\ 1
\end{array}\right)_l
\otimes\left(\begin{array}{c} 1\\ \epsilon
\end{array}\right)_r \sqrt{\Gamma_L}e^{-\mathrm{i} (m_L-\frac{\mathrm{i}}{2}\Gamma_L)
t_l}\cdot  \sqrt{\Gamma_S}e^{-\mathrm{i} (m_S-\frac{\mathrm{i}}{2}\Gamma_S)
t_r}
\bigg{)}
\label{phinew}\end{eqnarray}

Projecting (\ref{phinew}) onto $\left(\begin{array}{c} 1\\ 0
\end{array}\right)_l\otimes 
\left(\begin{array}{c} 1 \\ 0
\end{array}\right)_r $ to evaluate the amplitude $\psi_{11}$ associated to the decay intensity for the process $\phi\rightarrow ({\rm 2 neutral kaons})\rightarrow  \pi^+\pi^-$ we get

\begin{equation}\psi_{11}={1+|\epsilon|^2\over \sqrt 2(1-\epsilon^2)}\epsilon\sqrt{\Gamma_S\Gamma_L}\bigg{(}e^{-\mathrm{i} (m_S-\frac{\mathrm{i}}{2}\Gamma_S)
t_l}\cdot  e^{-\mathrm{i} (m_L-\frac{\mathrm{i}}{2}\Gamma_L)t_l}-
e^{-\mathrm{i} (m_L-\frac{\mathrm{i}}{2}\Gamma_L)
t_l}\cdot e^{-\mathrm{i} (m_S-\frac{\mathrm{i}}{2}\Gamma_S)
t_r}
\bigg{)}
\end{equation}

The previous reasoning can be generalised in order to establish that, in the Wave Function or Time Operator approach the joined-probability of firing of left and right detectors in the CP=+1 sector, which is equal to the modulus squared of $\psi_{11}$, is proportional, up to a constant in time factor, to the quantity $P^{11}_S(t_l,t_r)$.

This means that in the case that the survival probability and the corresponding pdf are proportional to each other it is not possible to discriminate the standard and non-standard approaches.

As we mentioned in the footnote \ref{footnoteremark} , in the case of the singlet state, a remarkable coincidence occurs so that the ratio between $(\frac{\partial }{\partial t_l}+\frac{\partial }{\partial t_r})P^{11}_S(t_l,t_r)$ and $ P^{11}_S(t_l,t_r)$ is a constant factor that does not depend on time.
Then the modulus squared of $\psi_{11}$ is proportional to the corresponding quantity evaluated in the standard approach, which is the pdf $P^{11}_S(t_l,t_r)$. So, the standard and non-standard approaches converge in their predictions regarding the singlet state (which was for instance produced in $\phi$ factories in Frascati).

 In the next section we shall show that it is not so for all entangled states, which opens the way to a new class of crucial experiments aimed at discriminating standard and non-standard predictions.

Curiously, in the literature relative to the study of the correlations between entangled mesons \cite{bertlmann,frascati,PRA63} it is often the analog of expression $ P^{11}_S(t_l,t_r)$ that is used in order to evaluate the joined-probabilities of detections {\it by unit of time} (the pdf quantities). For instance, in the reference \cite{frascati} the expression $P^{11}_S(t_l,t_r)$ is used in order to describe the joined pdf related to pion pair detections. This is somewhat disturbing because although it is appropriate to make use of such expressions in the framework of the Time Operator approach, it is not necessarily consistent to make use of them in the standard approach, in particular when {\it passive} detections are made. To our knowledge nobody makes use of the quantity $p^{11}_d=(\frac{\partial }{\partial t_l}+\frac{\partial }{\partial t_r})P^{11}_S(t_l,t_r)$ that we considered above, which is in our eyes the correct expression in the framework of the standard approach. This reveals, according to us, some confusion regarding the status of time in decay processes (that we already mentioned in section \ref{xxx} and also discuss in the appendix 3 where we compare the predictions made for passive continuous measurements with those made for active instantaneous ones). 
 
\subsection{Generalisation to other entangled states.\label{nosinglet}}

Let us now assume that kaon pairs are produced in the  EPR-Bohm $\psi^{\alpha}$ state defined by
 
 \begin{eqnarray}|\psi^{\alpha}\rangle
 ={1+|\epsilon|^2\over \sqrt 2(1-\epsilon^2)}\bigg{(}| \mathrm{K}_L\rangle_l \mathrm{K}_S\rangle_r - e^{i\alpha}| \mathrm{K}_S\rangle_l | \mathrm{K}_L\rangle_r\bigg{)},\label{phi}\end{eqnarray}

then, following the same of reasoning as in the previous section we conclude that the joined-probability $ p^{11\alpha}_d(t_l,t_r)$ of joined decay in the CP=+1 channel obeys

\begin{eqnarray} p^{11\alpha}_d(t_l,t_r)=(\frac{\partial }{\partial t_l}+\frac{\partial }{\partial t_r}) P^{11\alpha}_S(t_l,t_r)=\nonumber\\ (\frac{\partial }{\partial t_l}+\frac{\partial }{\partial t_r}){|\epsilon|^2\over  2(1-\epsilon^2)^2} 
|e^{-\mathrm{i} (m_L-\frac{\mathrm{i}}{2}\Gamma_L)
t_l}e^{-\mathrm{i} (m_S-\frac{\mathrm{i}}{2}\Gamma_S)
t_r} -
e^{i\alpha}e^{-\mathrm{i} (m_S-\frac{\mathrm{i}}{2}\Gamma_S)
t_l} e^{-\mathrm{i} (m_L-\frac{\mathrm{i}}{2}\Gamma_L)
t_r}|^2
\nonumber\\ \approx (\frac{\partial }{\partial t_l}+\frac{\partial }{\partial t_r})|\epsilon|^2 \cdot\bigg{(}
e^{-\Gamma_Lt_l-\Gamma_S t_r} +e^{-\Gamma_St_l-\Gamma_L t_r}-2e^{-\frac{(\Gamma_S+\Gamma_L)(t_l+t_r)}{2}}
cos (\Delta m (t_l-t_r)+\alpha)\bigg{)}
\nonumber\\ = (\Gamma_S+\Gamma_L)|\epsilon|^2 \cdot\bigg{(}
e^{-\Gamma_Lt_l-\Gamma_S t_r} +e^{-\Gamma_St_l-\Gamma_L t_r}-2e^{-\frac{(\Gamma_S+\Gamma_L)(t_l+t_r)}{2}}
cos (\Delta m (t_l-t_r)+\alpha)\bigg{)}
\nonumber\\ =(\Gamma_S+\Gamma_L)|\epsilon|^2|e^{-\mathrm{i} (m_L-\frac{\mathrm{i}}{2}\Gamma_L)
t_l}e^{-\mathrm{i} (m_S-\frac{\mathrm{i}}{2}\Gamma_S)
t_r} -
e^{i\alpha}e^{-\mathrm{i} (m_S-\frac{\mathrm{i}}{2}\Gamma_S)
t_l} e^{-\mathrm{i} (m_L-\frac{\mathrm{i}}{2}\Gamma_L)
t_r}|^2.\label{frascatipdtilde}\end{eqnarray}

The last equality shows that there exists an infinity of entangled states for which the standard and non-standard predictions coincide (up to a global calibration factor\footnote{It is straightforward to check that the predictions of the aforementioned ``hybrid'' approach also coincide with the standard and TWF-O predictions in this case but we consider that the hybrid approach is inconsistent, in particular in the case of passive measurements as we explain in appendix 3.}). In particular this is so for the singlet state ${1+|\epsilon|^2\over \sqrt 2(1-\epsilon^2)}\bigg{(}| \mathrm{K}_L\rangle_l \mathrm{K}_S\rangle_r - | \mathrm{K}_S\rangle_l | \mathrm{K}_L\rangle_r\bigg{)}$ but also for another Bell state, the state ${1+|\epsilon|^2\over \sqrt 2(1-\epsilon^2)}\bigg{(}| \mathrm{K}_L\rangle_l \mathrm{K}_S\rangle_r + | \mathrm{K}_S\rangle_l | \mathrm{K}_L\rangle_r\bigg{)}$. Now, this is not true for all entangled states. For instance, the states 
 \begin{eqnarray}|\psi^{\beta}\rangle
 ={1+|\epsilon|^2\over \sqrt 2(1-\epsilon^2)}\bigg{(}| \mathrm{K}_L\rangle_l \mathrm{K}_L\rangle_r - e^{i\beta}| \mathrm{K}_S\rangle_l | \mathrm{K}_S\rangle_r\bigg{)}\label{beta}\end{eqnarray}

are such that the joined-probability $ p^{11\beta}_d(t_l,t_r)$ of joined decay in the CP=+1 channel obeys

\begin{eqnarray} p^{11\beta}_d(t_l,t_r)=(\frac{\partial }{\partial t_l}+\frac{\partial }{\partial t_r}) P^{11\beta}_S(t_l,t_r)=\nonumber\\ (\frac{\partial }{\partial t_l}+\frac{\partial }{\partial t_r}){|\epsilon|^2\over  2(1-\epsilon^2)^2} 
|e^{-\mathrm{i} (m_L-\frac{\mathrm{i}}{2}\Gamma_L)
(t_l+
t_r)} -
e^{i\beta}e^{-\mathrm{i} (m_S-\frac{\mathrm{i}}{2}\Gamma_S)
(t_l+t_r)}|^2
\nonumber\\ \approx (\frac{\partial }{\partial t_l}+\frac{\partial }{\partial t_r})|\epsilon|^2 \cdot\bigg{(}
e^{-\Gamma_L(t_l+t_r)} +e^{-\Gamma_S(t_l+t_r)}-2e^{-\frac{(\Gamma_S+\Gamma_L)(t_l+t_r)}{2}}
cos (\Delta m (t_l+t_r)+\beta)\bigg{)}
\nonumber\\ = (\Gamma_S+\Gamma_L)|\epsilon|^2 \cdot\bigg{(}
e^{-\Gamma_Lt_l-\Gamma_S t_r} +e^{-\Gamma_St_l-\Gamma_L t_r}-2e^{-\frac{(\Gamma_S+\Gamma_L)(t_l+t_r)}{2}}
[cos (\Delta m (t_l+t_r)+\beta)\nonumber\\-{2\Delta m \over \Gamma_S+\Gamma_L}sin  (\Delta m (t_l+t_r)+\beta)   ]                         \bigg{)}
\nonumber\\ \not=(\Gamma_S+\Gamma_L)|\epsilon|^2|e^{-\mathrm{i} (m_L-\frac{\mathrm{i}}{2}\Gamma_L)
(t_l+
t_r)} -
e^{i\beta}e^{-\mathrm{i} (m_S-\frac{\mathrm{i}}{2}\Gamma_S)
(t_l+t_r)}|^2.\nonumber\\ =(\Gamma_S+\Gamma_L)|\epsilon|^2 \cdot\bigg{(}
e^{-\Gamma_L(t_l+t_r)} +e^{-\Gamma_S(t_l+t_r)}-2e^{-\frac{(\Gamma_S+\Gamma_L)(t_l+t_r)}{2}}
cos (\Delta m (t_l+t_r)+\beta)\bigg{)}\label{frascatibeta}\end{eqnarray}

This second class of maximally entangled states for which standard and non-standard predictions obviously differ comprises the two remaining Bell states

 ${1+|\epsilon|^2\over \sqrt 2(1-\epsilon^2)}\bigg{(}| \mathrm{K}_L\rangle_l \mathrm{K}_L\rangle_r +| \mathrm{K}_S\rangle_l | \mathrm{K}_S\rangle_r\bigg{)}$ and ${1+|\epsilon|^2\over \sqrt 2(1-\epsilon^2)}\bigg{(}| \mathrm{K}_L\rangle_l \mathrm{K}_L\rangle_r -| \mathrm{K}_S\rangle_l | \mathrm{K}_S\rangle_r\bigg{)}$.

\section{Discussion and conclusions.}
\subsection{Experimental proposal.}
As we have discussed thorughout our paper, some ambiguity remains present regarding the status of the statistical distributions of decay times. What are the experimentators exactly measuring? Integrated survival probabilites? Decay rates? Or even some intermediate quantity? Is it still conceivable to accept the existence of a Time Operator? or even of a Time Super Operator\footnote{The last possibility has been discussed in reference \cite{09superop} where we have shown that in the formalism of the Time Super Operator one can also derive non-standard predictions that could be tested by carefully repeating the Cronin and Fitch experiment.}?

As we have shown, the different alternatives lead to different experimentally testable predictions so that, in last resort, experiment ought to judge which is the most correct estimation. 

A possible way to test whether our ideas are relevant or not would be (section \ref{single}) to reproduce the original CP violation experiment \cite{perkins,christ}, but this time to carefully measure the decay rate with good enough resolution (which means over times shorter than the typical decay times involved) and in different time regimes ((a) of the order of $\tau_S$, (b) of the order of $\tau_L$, and (c) in-between where interferences are likely to occur). If this is done over several ranges of typical times (from $\tau_S$ to $\tau_L$ and in between), with an accurate calibration of the measured decay rates, then it is indeed possible in principle to discriminate between the predictions (\ref{superstandard}), (\ref{hybrid}) and (\ref{timeoperator}).

Another possibility is to prepare entangled states for which the ratio ${(\frac{\partial }{\partial t_l}+\frac{\partial }{\partial t_r})P^{11}_S(t_l,t_r)\over P^{11}_S(t_l,t_r)}$ is not a constant-in-time factor (in accordance with footnote \ref{footnoteremark} and section \ref{nosinglet}).

All this could also be done with any type of CP violating meson, for instance with $B$ particles in a set up similar to the one that exists in Tsukuba \cite{tsukuba}. For $B$ and $D$ particles the ``short'' and ``long'' lifetimes are very close to each other, but otherwise, all the machinery that we presented in the present paper can be applied at once.

It is our hope that this fine study will tell us which theory provides the best fit to reality and shed a new light on the status of Time in the Quantum Theory.

\subsection{Foundational issues.}
It is already difficult for the mind to represent itself the meaning of the usual wave function. As is well-known different interpretations exist regarding the sense to attribute to an amplitude of probability of presence in space. Needless to say that the situation does not improve if we seriously consider the possible existence of a wave function related to an amplitude of probability (of decay) in time. One of the counter-intuitive aspects of this (Temporal Wave Function or Time Operator) approach (which is also shared by the Super Time Operator approach) is that the statistical distribution of future decay times is ``contained'' in the state at time 0. It gives the uncomfortable feeling that the future is contained in the present and cannot be changed. This is misleading because actually if an experimentator actively chooses to change the environment (for instance by putting a slab of matter across the way of the particle\footnote{This corresponds to a measurement in a basis that diagonalizes the Strangeness operator.}
), the effective Hamiltonian changes accordingly.

Similar to what is done when light propagates along successive birefringent supports of which the major axes differ in orientations and absorption, one can adopt the TWF-O approach described in the present paper piecewise and impose requirements of continuity to pass from one segment to the next one. Of course it is a bit prematurate to consider these questions presently; the necessity to invoke our wave function model ought firstly to be motivated by convincing experiments.

It is also clearly out of the scope of the present paper to question the deep nature of CP-violation \cite{nature}.  A microscopic theory \cite{nobeljapan} of meson properties in the framework of chromodynamics has been succesfully tested in several experiments in the past and it is not our goal to challenge this approach. 

In this paper, we focus on a very particular aspect of meson phenomenology that concerns the status of Time and we remain confined to this particular domain. Actually, the fundamental question that we address
here essentially concerns the validity of the Time Operator description of the decay process, the existence of which has been the object of a long controversy in the past, with the hope that our experimental proposals constitute crucial tests that can bring an answer to this fundamental question.
We consider that CP-phenomenology is interesting as far as it allows us to study complex temporal interferences. It would be interesting to investigate how similar crucial experiments could be conceived and realized in low and intermediate energy physics, in the domain of Quantum Optics for instance\footnote{The time-bin approach that happened to be very useful in Quantum Cryptography in the past also leads to questions similar to the ones that were addressed in the present paper, for instance in the case that a particle is a superposition of two packets present in a same place at different times \cite{timebin,timebin2}. More recently, very impressive cavity QED experiments allowed to directly watch quantum jumps and also addressed the quantum Zeno paradox \cite{harochezeno}. At the end of appendix 3 (footnote \ref{cat}) we briefly mention an experiment that could be realised for measuring temporal interference between two cavity QED Schr\"odinger cat states.}.

Another interesting idea that has been suggested by our analysis of the measurement problem, and by the distinction between instantaneous and continuous decay is that an instantaneous measurement is nothing else than an idealized, sometimes convenient but illusory, approximation. As we discuss in the appendix 2, this approach makes it possible to circumvent the quantum Zeno paradox \cite{misrasud,chiuchiu}.

Last but not least, decay processes are an explicit manifestation of the Arrow of Time. It is not by chance that radio-active decay plays a prior role in the context of historical datation. $CP$ violation is also suspected to be at the origin of the dissymetry between matter and anti-matter in our universe, and it is not clear yet what is the deep nature of this symmetry-breaking. It is our hope that a fine study of temporal interference processes will contribute to a better understanding of these fundamental problems.



\section*{Appendix 1: Pauli's objection about the Time Operator.}
The possibility of defining a Time Operator in quantum mechanics is a controversial question. Pauli showed thanks to very simple arguments that if one could find an operator $\hat{T}$ that satisfies canonical commutation rules $[\hat{T}, \hat{H}]=i\hbar$ with the Hamiltonian operator $\hat{H}$ of a quantum system, then the spectrum of $\hat{H}$ ought to be unbounded by below, which clearly constitutes a physical impossibility.
A possible way to answer Pauli's objection is to define a ''super'' time operator that acts onto density matrices rather than onto pure states. It is not our purpose to investigate this question in the present work, although our present results are to a large extent inspired by a previous study of the time Superoperator \cite{09superop}. The idea that underlies the SuperTime Operator approach \cite{cour80,MPC,karpov1,cs,09superop} is that Pauli's objections are valid in the Hilbert space of pure states, but are not valid in the superoperator space, which is the space of linear operators acting on the space of pure states. Here we give a broader meaning to the Time Operator than the one given by Pauli: in our approach a Time Operator exists whenever the decay probability density function or decay rate can be expressed as the modulus square of a temporal amplitude $\psi(t)$. It is not necessary according to us that there exists an operator $\hat{T}$ that satisfies canonical commutation rules with the Hamiltonian operator in order that the Time Operator approach is valid. 

Actually it is easy to get confused by the literature of high energy physics because a very common technique that is used in order to estimate the mass and decay rate of an unstable particle (its complex energy) consists of measuring the distribution in energy of its decay products in order to fit it with a Breit-Wigner (or Lorentzian or Cauchy) distribution. The characteristic function of the Breit-Wigner distribution being equal to $C. e^{-\mathrm{i}(mt-\frac{\mathrm{i}}{2}\Gamma |t|) }$, this gives the feeling that high energy physicists follow the non-standard Time Operator approach but the situation is more subtle than that.

Indeed, let us assume \cite{girad} that the survival probability of an unstable particle obeys $P_s(t)=|<\psi(0),\psi(t)>|^2$ and that at time 0 the amplitude of probability that the particle energy is $E$ equals $\tilde{\psi}(E)$. Then, developing the wave function in the energy eigenbasis it is straightforward to show that $P_s(t)=|<\psi(0),\psi(t)>|^2=|\int_{E_{min.}}^{E_{max.}}dE e^{-i{Et\over \hbar}}|\tilde{\psi}(E)|^2|^2$, where the spectrum of the Hamiltonian runs from $E_{min.}$ to $E_{max.}$. It is impossible for the energy to take values that do not belong to the spectrum of the Hamiltonian so that $|\int_{E_{min.}}^{E_{max.}}dE e^{-i{Et\over \hbar}}|\tilde{\psi}(E)|^2|$ is close to $|\int_{-\infty}^{+\infty}dE e^{-i{Et\over \hbar}}|\tilde{\psi}(E)|^2|$ and the survival probability is equal to the squared modulus of the Fourier transform of the energy distribution.

Some problems occur in this approach; for instance it is not clear how to interpret the survival probability for negative times in the case of a Breit-Wigner (Lorentzian) distribution. Moreover, in the case that we impose a cutoff on the energy spectrum small discrepancies are likely to occur from the exponential behaviour. Now, they occur either for very short times (the Zeno regime) or very long times which are anyhow deprived of any significant statistical weight so that these corrections are out of reach of experimental observations \cite{girad}.

If instead we adopt in the present case, the non-standard, Time Operator Approach, we find \cite{09superop}(also assuming that the spectrum of the Hamiltonian is unbounded by below, which is a valid approximation in the Wigner-Weisskopf regime, in order to bypass Pauli's objection) that ${-d P_s(t)\over dt}=\Gamma\theta(t)e^{-\Gamma t}$=$|\Psi(t)|^2$=$|\sqrt{\Gamma}\theta(t)e^{-\mathrm{i}(m-\frac{\mathrm{i}}{2}\Gamma)t ) }|^2$=$|\int_{-\infty}^{+\infty}dE e^{-i{Et\over \hbar}}\hat{\Psi}(E)|^2$ where $\theta(t)$ is the Heaviside function and $\hat{\psi}(E)$=${1\over 2\pi}\int_{-\infty}^{+\infty}dt e^{-i{Et\over \hbar}}\Psi(t)$=$-i\sqrt{{\Gamma\over 2\pi}}{1\over (E-(m-\frac{\mathrm{i}}{2}\Gamma))} $.

Incidentally, the modulus squared of $\hat{\psi}(E)$ is precisely equal to the Breit-Wigner distribution:

$|\hat{\psi}(E)|^2$=$N.{1\over (E-m)^2+(\Gamma/2)^2}$ with $N$ a normalisation factor.

Therefore it is not possible on the basis of exponential decay to discriminate the standard and non-standard approaches (as we also noted in section \ref{wavefuc} following a different reasoning).

Besides, we consider that the assumption made above that the survival probability obeys $P_s(t)=|<\psi(0),\psi(t)>|^2$ is not universally valid. For instance when the energy spectrum contains a discrete part (a situation that occurs in kaon phenomenology but also in quantum optics for instance when QED cavities are considered), so that Rabi oscillations and decay processes simultaneously occur it is no longer guaranteed that the function $|<\psi(0),\psi(t)>|^2$ is monotonously decreasing (see also footnote \ref{monotonous}) in which case it is clearly not correct to assume that the decay probability $P_s(t)$ is equal to the auto-correlation of the wave function $|<\psi(0),\psi(t)>|^2$.

It is on the basis of these remarks that we adopted the very general {\it definitions} (\ref{exp}) and (\ref{exprategen}) that in our eyes encapsulate the quintessence of the standard and non-standard (Temporal Wave Function or Time Operator) approaches.

As we have shown in the paper, in the case that the decay departs from the most common (exponential) behaviour, for instance when two exponential decay processes coherently superpose to each other, the standard and non-standard approach can be discriminated experimentally.

A simple way to conceive the difference between both approaches is to consider that in the standard approach the superposition principle concerns ponctual times while in the non-standard approach it concerns time intervals or durations.

\section*{Appendix 2: Circumventing the Zeno paradox.}
The aim of this section is two-fold:

1) we shall show that instantaneous measurements (that are seen as an idealization of very short measurements) are not incompatible with the approach presented in the section \ref{sexion} and that in principle they can be realized during the completion of continuous measurements without affecting the statistical distribution of their outcomes. 

2) consequently, the Zeno paradox \cite{misrasud,chiuchiu} according to which a series of instantaneous measurements freezes the state of the system under measurement (in the limit that the delay between two successive instantaneous measurements becomes infinitely short) loses its meaning: in our case, it is harmless to carry out such measurements, and the distinction between instantaneous and continuous measurements loses its sense\footnote{This is true in the case that the formalism presented in the section \ref{sexion} is valid which implies that the decay is exponential¤. Zeno-related effects have been observed experimentally \cite{zenoexp} which shows that this approach is not universally valid.}.

Let us consider gedanken-experiments where certain measurements are assumed to be instantaneous as is done for instance in ref.\cite{bertlmann}. If we come back to the analogy with polarised light that was evocated at the beginning of section \ref{standard}, an instantaneous measurement is a realistic possibility in that case. Indeed, one could in principle measure the polarisation of a monochromatic transverse electric field (of pulsation $\omega$) in a birefringent guide by separating two orthogonal polarisation components with a polarizing beamsplitter, performing a non demolitive measure of the polarisation and reinjecting the photons in the fibre. In principle this process can be realized in a very short time which makes it a good candidate for an instantaneous measurement.

If we would realize the counterpart of such an experiment at the level of kaons, measuring the CP observable instead of photonic polarisation, the instantaneous probability to observe a pair of pions at time $t$ would be equal to the product of the probability that the particle survives by then (which is ${\psi_1^*(t)\psi_1(t)+\psi_2^*(t)\psi_2(t)\over \psi_1^*(0)\psi_1(0)+\psi_2^*(0)\psi_2(0)}$) with the probability of observing the CP=+1 component at that time (which is ${\psi_1^*(t)\psi_1(t)\over \psi_1^*(t)\psi_1(t)+\psi_2^*(t)\psi_2(t)}$). The resulting probability $P^{+1}(t)$  to observe CP=+1 during an hypothetical instantaneous CP measurement at time $t$ would then obey

\begin{equation}\label{uninstant}P^{+1}(t)={\psi_1^*(t)\psi_1(t)\over \psi_1^*(0)\psi_1(0)+\psi_2^*(0)\psi_2(0)}.\end{equation}

More generally, in the case that we interrupt the ''free''evolution (\ref{fh1}) by an ''instantaneous'' measurement in the basis $\{|+\rangle$, $|-\rangle\}$ of our choice, the probability to observe the result + (-) would be  (as is also explained in \cite{bertlmann}) equal to 

$|\psi_1(t)\langle +|\mathrm{K}_1\rangle + \psi_2(t)\langle +|\mathrm{K}_2\rangle|^2$ ($|\psi_1(t)\langle -|\mathrm{K}_1\rangle + \psi_2(t)\langle -|\mathrm{K}_2\rangle|^2$).

Let us assume now for a while that instantaneous measurements are possible and that we interrupt the ''free''evolution (\ref{fh1}) by an ''instantaneous'' measurement in the CP basis. We shall now show that (in the absence of CP violation) this interruption is not likely to lead to Zeno-like effects so that both pictures, the continuous and instantaneous measurement pictures peacefully coexist\footnote{We refer here to the so-called quantum Zeno paradox according to which a series of repeated measurements performed on a two-level sytem freezes the coherent Rabi oscillations of the system, in the limit that the measurements are repeated quickly enough. We shall adopt a very broad acceptance of the Zeno paradox in the sense that, in what follows, we shall say that Zeno-like effects are present whenever an observation influences in average the statistics of subsequent measurements.}. There exist several ways to understand why the quantum Zeno paradox is not relevant in the present case.

Firstly, we are not free to monitor the evolution of the system, so that there is no Rabi oscillation that could interfere with the (self) measurement process and open the door to Zeno-like effects.

Secondly, even if we would have the possibility to realize instantaneous measurements in the  CP-eigenbasis $\{|\mathrm{K}_1\rangle$, $|\mathrm{K}_2\rangle\}$, separated by an evolution that would obey (\ref{fh1}), this would not affect the statistics of the outcomes. Indeed, assuming that an instantaneous measurement is performed at time $t_a$ it will collapse the state  $\left(\begin{array}{c}
\psi_1(t_a)\\ \psi_2(t_a)
\end{array}\right)$ onto $|\mathrm{K}_1\rangle=\left(\begin{array}{c}
1\\0
\end{array}\right)$ with probability $|\psi_1(t_a)|^2$ and onto $|\mathrm{K}_2\rangle=\left(\begin{array}{c}
0\\1
\end{array}\right)$ with probability $|\psi_2(t_a)|^2$. If now a second instantaneous measurement is performed at time $t_b$ it will collapse the state  $|\mathrm{K}_1\rangle$ onto itself with certainty 1 in the case that it survives, which occurs with probability $P_s(t_b-t_a)={|\psi_1(t_b)|^2\over |\psi_1(t_a)|^2} $ and $|\mathrm{K}_2\rangle$ onto itself with probability $P_s(t_b-t_a)={|\psi_2(t_b)|^2\over |\psi_2(t_a)|^2}$, where we made use of Eqn.(\ref{exp}) and of the properties of the exponential distribution.

At the end of the day, the probabilities to measure a CP=+1 or CP=-1 eigenvalue at time $t_b$ are respectively equal to  $|\psi_1(t_a)|^2\cdot {|\psi_1(t_b)|^2\over |\psi_1(t_a)|^2}=|\psi_1(t_b)|^2$ ($|\psi_2(t_a)|^2\cdot {|\psi_2(t_b)|^2\over |\psi_2(t_a)|^2}=|\psi_2(t_b)|^2$). 

These probabilities are exactly the same whether an instantaneous CP measurement is performed at time $t_a$ or not. 

In virtue of the expressions  (\ref{exp}-\ref{exp2}) this remarkable property is also valid that we consider the projected survival probabilities or the cumulated survival probabilities:

\begin{equation}P_s^i(t_b)={\psi_i^*(t_b)\psi_i(t_b)\over \psi_i^*(0)\psi_i(0)}={\psi_i^*(t_b)\psi_i(t_b)\over \psi_i^*(t_a)\psi_i(t_a)}\cdot{\psi_i^*(t_a)\psi_i(t_a)\over \psi_i^*(0)\psi_i(0)}.\end{equation}

and

\begin{equation}P_s(t_b)={\|\psi(t_b)\|^2\over \|\psi(0)\|^2}={\|\psi(t_b)\|^2\over \|\psi(t_a)\|^2}\cdot{\|\psi(t_a)\|^2\over \|\psi(0)\|^2}.\end{equation}

Now, it is worth noting that this is true as long as the two channels (1 and 2) are decoupled, as long as the decay is exponential in each channel, and as long as the instantaneous measurement basis is the CP eigen basis.

\section*{Appendix 3. About the polarisation measurement analogy.}
\subsection*{EPR-Bohm kaonic states.}

In ref.\cite{frascati}, it is assumed that the joint-probability of detecting a pair of pions  $\pi^+\pi^-$ at time $t_l$ in the left region and $t_r$ in the right one (called the decay intensity for the process $\phi\rightarrow ({\rm 2 neutral kaons})\rightarrow  \pi^+\pi^-$) is equal to the modulus square of the {\it ``amplitude''}

 $\langle \pi^+\pi^- | \mathrm{K}_1\rangle_l \langle \pi^+\pi^- | \mathrm{K}_1\rangle_r  \  _l\langle \mathrm{K}_1| _r\langle \mathrm{K}_1|\psi(t_l,t_r)\rangle$.

This hybrid expression contains an in-product between states that belong to the same Hilbert space ($| \mathrm{K}_1\rangle_l| \mathrm{K}_1\rangle_r$ and $|\psi(t_l,t_r)\rangle$) as well as an {\it in-product} between states that do not belong to the same Hilbert space ($| \pi^+\pi^-\rangle$ and $|\mathrm{K}_1\rangle$). Intuitively, the last expression ought to be interpreted as an amplitude of probability of transition {\it by unit of time}, although this point is not made explicit in ref.\cite{frascati}.

Making use of the expression (\ref{phi}) of the entangled state $|\psi(t_l,t_r)\rangle$ this amplitude is expressed as 

\begin{eqnarray}{1+|\epsilon|^2\over \sqrt 2(1-\epsilon^2)} \bigg{(}\langle \pi^+\pi^- | \mathrm{K}_1\rangle_l  \langle \mathrm{K}_1|\mathrm{K}_L\rangle_l
\langle \pi^+\pi^-|\mathrm{K}_1\rangle_r  \langle\mathrm{K}_1|{\mathrm{K}}_S\rangle_r
e^{-\mathrm{i} (m_L-\frac{\mathrm{i}}{2}\Gamma_L)
t_l}e^{-\mathrm{i} (m_S-\frac{\mathrm{i}}{2}\Gamma_S)
t_r} \nonumber\\
-\langle \pi^+\pi^- | \mathrm{K}_1\rangle_l  \langle \mathrm{K}_1|\mathrm{K}_S\rangle_l
\langle \pi^+\pi^-|\mathrm{K}_1\rangle_r  \langle\mathrm{K}_1|{\mathrm{K}}_L\rangle_re^{-\mathrm{i} (m_S-\frac{\mathrm{i}}{2}\Gamma_S)
t_l} e^{-\mathrm{i} (m_L-\frac{\mathrm{i}}{2}\Gamma_L)
t_r}\bigg{)}\end{eqnarray}

So that the decay intensity for the process $\phi\rightarrow ({\rm 2 neutral kaons})\rightarrow ({\rm 2 pion pairs} \pi^+\pi^-)$ is equal to

\begin{eqnarray}{|\epsilon|^2\over  2(1-\epsilon^2)^2} |\langle \pi^+\pi^- | \mathrm{K}_1\rangle|^4  |
e^{-\mathrm{i} (m_L-\frac{\mathrm{i}}{2}\Gamma_L)
t_l}e^{-\mathrm{i} (m_S-\frac{\mathrm{i}}{2}\Gamma_S)
t_r} -
e^{-\mathrm{i} (m_S-\frac{\mathrm{i}}{2}\Gamma_S)
t_l} e^{-\mathrm{i} (m_L-\frac{\mathrm{i}}{2}\Gamma_L)
t_r}|^2\nonumber\\ \approx |\epsilon|^2 |\langle \pi^+\pi^- | \mathrm{K}_S\rangle|^4  \cdot\bigg{(}
e^{-\Gamma_Lt_l-\Gamma_S t_r} +e^{-\Gamma_St_l-\Gamma_L t_r}
-2e^{-\frac{(\Gamma_S+\Gamma_L)(t_l+t_r)}{2}}
cos (\Delta m (t_l-t_r))\bigg{)},\label{frascati}
\end{eqnarray}
in conformity with expression (2) of ref.\cite{frascati}.

Let us assume that by a magical intervention we project at time $t_l$ in the left region and $t_r$ in the right region the pair of kaons onto a product-basis of CP-eigenstates and that, after this, we passively wait that the pair of kaons decay, and we count the number of pairs of pions  $\pi^+\pi^-$ in each region.

It is easy to check that then the probability of joint-firing of the detectors is equal to the product of the probability that the pair survives times the probability that each kaon is projected onto the CP=+1 eigenspace times the probability that a $| \mathrm{K}_1\rangle$ state decays (by unit of time) onto a pion pair, which is $\bigg{(}
{e^{-\Gamma_Lt_l-\Gamma_S t_r} +e^{-\Gamma_St_l-\Gamma_L t_r}
\over 2}\bigg{)}$ times the decay intensity for the process $\phi\rightarrow ({\rm 2 neutral kaons})\rightarrow  \pi^+\pi^-$) that can be found in Eqn.\ref{frascati} in conformity with expression (2) of ref.\cite{frascati}.

 Conditioning on the survival of the pair at times $t_l$ in the left region and $t_r$ in the right region resorts to divide the probability of joint-firing by the normalisation factor $\bigg{(}
{e^{-\Gamma_Lt_l-\Gamma_S t_r} +e^{-\Gamma_St_l-\Gamma_L t_r}
\over 2}\bigg{)}$ so that we find a full agreement with the expression (2) of ref.\cite{frascati}.

\subsection*{One particle case-the ``hybrid'' approach.}

As we show below, by a hybrid reasoning similar to the reasoning explained in the previous section, we are also able to reproduce the interpretation of the physical results adopted during the Cronin and Fitch experiment. We checked that several formulas that appear in the literature \cite{bertlmann,perkins,frascati,PRA63} can also derived by following this type of reasoning which constitutes in our eyes a hybrid reasoning similar to the one examplified at the level of eqn.\ref{hybrid}.

This reasoning is hybrid in the sense that it can be justified in the case that an instantaneous measurement is performed but is difficult to justify when the measurement is passive and continuous in time.
According to us, this is representative of the widespread confusion that prevails regarding the distinction between survival probabilites and temporal pdf.

We are convinced that this distinction is important: an instant of time should not be confused with an interval of time. As we have shown in appendix 2, there exists, in first approximation, a situation of peaceful coexistence between instantaneous measurements and continuous measurements, but as long as we consider somewhat complex situations, for instance those where temporal interference effects are present, it becomes very important to establish the necessary distinctions and to follow the correct reasonings. In last resort, experiments can indicate to us which are the right choices to perform.

What is certain is that it is potentially misleading to trust blindly the analogy between polarisation measurements and CP measurements because such analogies have a very limited relevance (as we mentioned briefly in the section \ref{contmeas}). Essentially this is due to the fact that the former ones are active and nearly instantaneous processes, while measuring CP by analysing the decay products is a passive and continuous in time process\footnote{\label{fine}It is worth noting that in the case that the experimentalist actively chooses to interpose a slab of matter across the trajectory of the incoming kaon, it is fully consistent to treat the measurement process as an instantaneous process and to resort to the analogy between quasi-spin and polarisation (see for instance section F in ref.\cite{PRA63}).}.

If for instance, resorting to this misleading analogy, we assume that, as happens in the Cronin and Fitch experiment, we are far enough from the source of $|\mathrm{K}^0\rangle$ states so that only the Long states $|\mathrm{K}_L\rangle$ $\approx \epsilon |\mathrm{K}^1\rangle$+$|\mathrm{K}^2\rangle$ survive and that by ``a magical intervention'' we instantaneously project this state onto the CP-eigenstates we find  
$|\mathrm{K}^1\rangle$ with probability close to $|\epsilon|^2$ and $|\mathrm{K}^2\rangle$ with probability close to 1. Now, $|\mathrm{K}^1\rangle$$\approx$ $|\mathrm{K}_S\rangle$ and $|\mathrm{K}^2\rangle$$\approx$ $|\mathrm{K}_L\rangle$ so that in such a case the ratio of the production rates of the pairs and triplets is no longer $|\epsilon|^2$ but rather ${\tau_1\over \tau_2}$$|\epsilon|^2$ =$({\Gamma_S\over \Gamma_L})^{-1}$$|\epsilon|^2$ because short states decay quite faster than long states. This could explain the appearance           of the mysterious correction factor ${\tau_1\over \tau_2}$ =$({\Gamma_S\over \Gamma_L})^{-1}$ that we mentioned at the end of section \ref{xxx}.

Indeed, in the case that a state $\alpha  |\mathrm{K}^1\rangle$+$\beta |\mathrm{K}^2\rangle$ is replaced by an incoherent mixture $\rho=|\alpha|^2 |\mathrm{K}^1\rangle\langle\mathrm{K}^1|+|\beta|^2 |\mathrm{K}^2\rangle\langle\mathrm{K}^2|$, the pion pair (triplet) production is proportional in first approximation to $|\alpha|^2\Gamma_S$ ($|\beta|^2\Gamma_L$) so that their ratio is equal to ${|\alpha|^2\Gamma_S\over |\beta|^2\Gamma_L }$=${|\alpha|^2\over |\beta|^2 }{\Gamma_S\over\Gamma_L }$.

In our eyes, nevertheless, this reasoning is not valid because it is incorrect to neglect the coherence between the two CP channels. In quantum optics for instance it is well-known that dark states  \cite{scully} can be generated by preparing coherent superpositions of energy states in such a way that the resulting amplitude of transition towards the final state vanishes. Similarly, temporal quantum beats have been observed in the case of $V$ transitions (section 1.4 of ref.\cite{scully}) for which coherence is present\footnote{\label{cat}Similarly, in the case that Schr\"odinger cat states are created in QED cavities by coherently superposing coherent states with opposite phases \cite{harochecat,haroche}, coherence survives for a while provided the number of photons in the cat states is low enough. In principle quantum beats are likely to occur when the first photon is emitted, before the decoherence process becomes effective.}. Having this example in mind, it is obvious to understand why it is incorrect to treat the state $|\mathrm{K}_L\rangle$ as an incoherent mixture of its CP components: the reason is that this cannot be done without altering its decay behaviour.

That this hybrid approach is inconsistent can be seen for instance by comparing the full decay rate, cumulated over the CP=+1 and -1 sectors after having prepared at time 0 a state $|\mathrm{K}^0\rangle$.

Indeed, in the same way that we drive the expression (\ref{intens}) from (\ref{scramble}), we find, projecting now onto the CP=-1 sector
\begin{eqnarray}P^2_s(t)&=&P^2_s(0)\cdot{\psi_2^*(t)\psi_2(t)\over \psi_2^*(0)\psi_2(0)}\nonumber\\
 &=&{P^2_s(0)\over |1+\epsilon|^2}\,\bigg{(}|\epsilon |^2e^{-\Gamma_S t}+
e^{-\Gamma_L t}+
2|\epsilon |e^{-({\Gamma_S +\Gamma_L\over 2})t} \cos
\big{(}\triangle m t-\arg(\epsilon)\big{)}\bigg{)},
\label{intens2}\end{eqnarray}

In the case that we can consistently neglect $|\epsilon|$, $-{\partial |\psi_1(t)|^2\over \partial t}=\Gamma_S|\psi_1(t)|^2$ and ${-\partial |\psi_2(t)|^2\over \partial t}=\Gamma_L|\psi_2(t)|^2$ so that the hybrid approach is valid. Similarly the hybrid approach is justified whenever at time $t$ the experimentalist chooses to measure in the Strangeness eigenbasis by interposing a slab of matter along the kaon trajectories because this process kills the coherences in the strangeness basis (in accordance with our analysis of section \ref{sexion} and footnote \ref{fine}).

Now, in general, the hybrid approach is inconsistent as can be seen by considering the full decay rate:

\begin{equation}{\partial (|\psi_1(t)|^2+ |\psi_2(t)|^2)\over \partial t}\not=\Gamma_S|\psi_1(t)|^2+\Gamma_L|\psi_2(t)|^2,\end{equation}
due to the presence of CP-related effects so that the decay rate obtained by summing over the CP=+1 and -1 channels differs from the full decay rate, which is according to us an inconsistency.

It is only in the limit where we neglect CP violation ($\epsilon$=0) that both quantities are equal, in agreement with our analysis of section \ref{sexion}. Such an approximation is of course meaningless in the case that we are tracking CP violation effects.


\begin{thebibliography}{99}
\begin{footnotesize}
\bibitem{booktime}J.G.Muga, R.Sala Mayato and I.L.Egusquiza (eds) (2002), ``Time in Quantum Mechanics'', (Springer,
Berlin).
\bibitem{mielnik}B.Mielnik (1994): {\it ``The screen problem''}. Foundations of physics, {\bf 24}, 8, 1113-1129.
\bibitem{cushing}J.T. Cushing (1995):  {\it``Quantum tunneling times: a crucial test for the causal program?''}, Foundations of Physics \textbf{25}, 2, 269-280.
\bibitem{wignernewton}T. D. Newton and E. P. Wigner (1949):  {\it``Localized states for elementary systems''}, Rev. Mod. Phys. {\bf 21}, 400.
\bibitem{BG03} Bassi, A., Ghirardi, G.C. (2003):  {\it``Dynamical reduction models''},
  \textit{Phys. Rep.} \textbf{379}: 257-426.
\bibitem{EPR} A. Einstein, B. Podolsky, and N. Rosen (1935): \emph{Phys.
Rev.}, \textbf{47}, 777.
\bibitem{Bell} Bell, J. S. (1987): ``Speakable and unspeakable in
    quantum mechanics'', Cambridge: Cambridge University Press.
    \bibitem{busch}P. Busch, {\it``The time-energy uncertainty relation''}, in ``Time in Quantum Mechanics'', J.G.Muga, R.Sala Mayato and I.L.Egusquiza (eds) (2002),  (Springer,
Berlin).

    \bibitem{09superop} M. Courbage,  T. Durt, S.M. Saberi Fathi (2009):  {\it``A new formalism for the estimation of the CP-violation
parameters''}, arXiv:quant-ph/0907.2514v1.


\bibitem{09temporal} M. Courbage,  T. Durt, S.M. Saberi Fathi  (2009):  {\it``A wave-function model for the CP-violation in mesons''}, arXiv:quant-ph/0903.4143v1.
\bibitem{christ} J. H. Christenson, J. W. Cronin, V. L. Fitch and
R. Turlay  (1964):  {\it``Evidence for the $2\pi$ decay of the $\mathrm{K}_2$
meson'',} Phys. Rev. Lett. \textbf{13}, 138.
\textbf{63}, 54.
\bibitem{WW} V. Weisskopf, E. Wigner (1930):   {\it``Berechnung der naturlichen Linienbreite auf
Grund der Diracschen Lichttheorie''}, Zeitschrift f\''ur Physik.

\bibitem{gamow} G. Gamow (1928), Z. Phys. \textbf{51}, 537.


\bibitem{girad} L. Fonda, G. C. Ghirardi and A. Rimini  (1978):  {\it``Decay theory of
unstable quantum systems''}, Rep. Prog. Phys., \textbf{41} 589-630.

\bibitem{misrasud}  B. Misra, E.C.G. Sudarshan  (1977):  {\it``The Zeno's Paradox in Quantum Theory''},
J. Math Phys., \textbf{18}(4), 756.
\bibitem{chiuchiu} C.B. Chiu, E.C.G. Sudarshan and B. Misra (1977):  {\it``Time evolution of unstable quantum states and a resolution of Zeno's paradox''}, Phys.Rev. D {\bf 16}. 520. 
\bibitem{arrival}C. Anastopoulos (2008):  {\it``Time-of-arrival probabilities and quantum
measurements: III Decay of unstable states''}, J. Math. Phys. 49, 022103.


\bibitem{ghirardiEPR} G.C. Ghirardi, R. Grassi, A. Rimini and T. Weber (1988):  {\it``Experiments of the EPR type involving CP-violation do not allow faster-than-light communication between distant observers''}, Europhys. Lett. {\bf 6} (2), pp 95-100.

\bibitem{hokim} Q. Ho-Kim, X-Y Pham, (1998): ``Elementary
Particles and Their Interactions'', (Springer, Berlin-Heidelberg).
\bibitem{perkins}D. H. Perkins  (1987): ``Introduction to High Energy Physics'', Eds.
Addison-Wesley.


 \bibitem{Lee}T.D. Lee and C.S. Wu (1966): Annu. Rev. Nucl. Sci. {\bf 16}, 511.
\bibitem{Lipkin} H.J. Lipkin (1968): Phys. Rev. {\bf 176}, 1715.
\bibitem{bertlmann} Bertlmann R. (2006):  {\it``Entanglement, Bell inequalities and Decoherence in Particle Physics''},  in ``Quantum Coherence From Quarks to Solids'', Lecture Notes in Physics, Springer Berlin-Heidelberg, 1-45.


\bibitem{leebook}T. D. Lee (1981): ``Particle Physics and Introduction to Field
Theory'', Harwood Academic Publishers.

\bibitem{papaliolos}C. Papaliolos (1967):  {\it``Experimental test of a hidden variable quantum theory''}, Phys.Rev.Letters, {\bf 18}, n$^0$ 15, 622. 
\bibitem{bramon} A. Bramon, G. Garbarino, and B. Hiesmayr (2007):  {\it``Quantum mechanics with neutral kaons''}, Invited Talk at the Final EURIDICE Meeting 'Effective Theories of Colour and Flavour: from EURODAPHNE to EURIDICE, Kazimierz (Poland), August 24-27, 2006, hep-ph/0703152.
\bibitem{frascati}Ambrosino {\it et al.} (2006):  {\it``First observation of quantum interference in the process
$\phi\rightarrow K_SK_L \rightarrow \pi^+\pi^-\pi^+\pi^-$-: A test of quantum mechanics
and CPT symmetry''},  \emph{ Phys. Lett. B} {\bf 642} 315-321.
\bibitem{griffith} D. Griffiths (2008): ``Introduction to Elementary Particles'', {Eds. Wiley-VCH}.

\bibitem{PRA63} R. Bertlmann and B. Hiesmayr (2001):  {\it``Bell inequalities for entangled kaons and their unitary time evolution''}, Phys. Rev. A, vol.{\bf 63}, 062112.
\bibitem{iontraps} W. Paul, (1990):  {\it``Electromagnetic traps for charged and neutral particles''}, Rev. Mod. Phys. {\bf 62}, 531.
\bibitem{haroche} S. Haroche and J-M Raimond (2006):  ``Exploring the Quantum Atoms, Cavities and Photons'', Oxford University Press.
\bibitem{wineland} R.J. Raffac, B.C. Young, J.A. Young, W.M. Itano and J.C. Bergquist (2000):  {\it``Sub-dekahertz ultraviolet spectroscopy of $^{199}$Hg$^+$''}, Phys. Rev. Lett. {\bf 85}, 2462.
\bibitem{hansche} M. Niering, R. Holzwarth, J. Reichert, P. Pokasov, Th. Udem, M. Weitz, and T. W. HŠnsch; P. Lemonde, G. Santarelli, M. Abgrall, P. Laurent, C. Salomon and A. Clairon (2000):  {\it``Measurement of the Hydrogen 1S- 2S Transition Frequency by Phase Coherent Comparison with a Microwave Cesium Fountain Clock''}, Phys. Rev. Lett. {\bf 84}, 5496Ð5499.
\bibitem{fried} K. Friedrichs,  (1948):  {\it ``On the perturbation of continuous spectra''},
Communications on Appl. Math. \textbf{1}, 361-406.
\bibitem{CDS} M. Courbage,  T. Durt, S.M. Saberi Fathi (2007):
 {\it``Two-Level Friedrichs model and Kaonic phenomenology''},  Physics
letters A \textbf{362}, 100-104.
\bibitem{cds2} M. Courbage,  T. Durt, S.M. Saberi Fathi (2007): {\it``Quantum-mechanical
decay laws in the neutral Kaons''}, Journal of Physics A : Math.
Theor. \textbf{40}, 2773-2785.
\bibitem{bellstate} S.L. Braunstein, A. Mann and M. Revzen, {\it ``Maximal violation of Bell inequalities for mixed states''}, Phys. Rev. Lett., {\bf 68},
 (1992) 3259-3261.
\bibitem{cplear}Apostolakis {\it et al.} (1998): {\it``An EPR experiment testing the non-separability of the wave function''},  Phys. Lett. B, {\bf 422} 339-348.
\bibitem{tsukuba}A. Go {\it et al.} (2007): {\it``Measurement of EPR-type flavour entanglement in $\Upsilon$(4S) --- $>B^0$ anti-$B^0$ decays''}, By Belle Collaboration BELLE-2006-40, KEK-2006-61, Phys.Rev.Lett. {\bf 99}:131802. 
\bibitem{cour80} M. Courbage (1980): {\it``On necessary and sufficient conditions
for the existence of time and entropy operators''}, Lett. Math. Phys.
\textbf{4}, 425.
\bibitem{MPC} B. Misra, I. Prigogine and M. Courbage,  (1983): {\it``Lyapounov variables: entropy and measurement in quantum mechanics''}, in ''Quantum theory and measurement'' eds.
J.A. Wheeler and W.H. Zurek, Princeton, N-J, 687-693.

\bibitem{karpov1} Ordonez G Petrosky T, Karpov E, Prigogine
I . (2001): {\it``Explicit construction of a time superoperator for quantum
unstable systems''},  \emph{Chaos Solitons and Fractals }{\bf 12}
2591-2601.
\bibitem{cs} M. Courbage, S.M. Saberi Fathi; {\it``Decay  probability distribution of quantum-mechanical unstable
systems and  time operator''}  (2008):  Physica  A,  \textbf{387}, Issue 10,
1,  2205-2224.

\bibitem{zenoexp}M. C. Fischer, B. Guti\v{z}rrez-Medina, and M. G. Raizen: {\it ``Observation of the
 Quantum Zeno and Anti-Zeno Effects in an Unstable System''}, Phys. Rev. Lett. \textbf{87}, 040402
 (2001).
 \bibitem{nature} P. Ball (2009): {\it ``Quark statistics shed light on Universe's symmetry''}, Nature News, \textbf{458}, 559.
\bibitem{nobeljapan} http://nobelprize.org/nobel$_{-}$prizes/nobelguide.pdf
\bibitem{timebin} I. Marcikic, H. de Riedmatten, W. Tittel, V. Scarani, H. Zbinden and N. Gisin (2002): {\it``Femtosecond Time-Bin Entangled Qubits for Quantum Communication''}, Phys. Rev. A {\bf 66},062308.
\bibitem{timebin2}T. Durt and B. Nagler: {\it : ''Quantum cryptographic encryption in three complementary bases through a Mach-Zehnder set up'',} in the book ``Probing the structure of Quantum Mechanics:  nonlinearity, nonlocality, computation and axiomatics'', Editors. D. Aerts, M. Czachor and T. Durt, World Scientific, Singapore(2000)259-286.
\bibitem{harochezeno} J. Bernu, S. Del\'eglise, C. Sayrin, S. Kuhr, I. Dotsenko, M. Brune, J. M. Raimond and S. Haroche (2008): {\it``Freezing Coherent Field Growth in a Cavity by the Quantum Zeno Effect''}, Phys. Rev. Lett. {\bf 101},  180402.

\bibitem{scully} M.O. Scully and M.S. Zubairy (1997): {\it ``Quantum Optics''} Cambridge University Press.
\bibitem{harochecat} M. Brune, S. Haroche and J. M. Raimond; L. Davidovich and N. Zagury (1991):  {\it``Manipulation of photons in a cavity by dispersive atom-field coupling: Quantum-nondemolition measurements and generation of ÔÔSchr\"odinger catÕÕ states''}, Phys. Rev. A {\bf 45}, 5193Ð5214























\end{footnotesize}
\end{thebibliography}
\end{document}